\documentstyle[multicol,aps,epsf]{revtex}

\title{On the construction of high-order force gradient algorithms \\
       for integration of motion in classical and quantum systems}

\author{I. P. Omelyan,$^{1,2}$ I. M. Mryglod,$^{1,2}$ and R. Folk$^2$}

\address{$^1$Institute for Condensed Matter Physics,
         1 Svientsitskii Street, UA-79011 Lviv, Ukraine}
\address{$^2$Institute for Theoretical Physics, Linz University,
         A-4040 Linz, Austria}

\date{\today}

\begin{document}

\maketitle

\begin{abstract}

A consequent approach is proposed to construct symplectic force-gradient
algorithms of arbitrarily high orders in the time step for precise
integration of motion in classical and quantum mechanics simulations.
Within this approach the basic algorithms are first derived up to the
eighth order by direct decompositions of exponential propagators and
further collected using an advanced composition scheme to obtain the
algorithms of higher orders. Contrary to the scheme by Chin and Kidwell
[Phys. Rev. E {\bf 62}, 8746 (2000)], where high-order algorithms are
introduced by standard iterations of a force-gradient integrator of
order four, the present method allows to reduce the total number of
expensive force and its gradient evaluations to a minimum. At the
same time, the precision of the integration increases significantly,
especially with increasing the order of the generated schemes. The
algorithms are tested in molecular dynamics and celestial mechanics
simulations. It is shown, in particular, that the efficiency of the
new fourth-order-based algorithms is better approximately in factors
5 to 1000 for orders 4 to 12, respectively. The results corresponding
to sixth- and eighth-order-based composition schemes are also presented
up to the sixteenth order. For orders 14 and 16, such highly precise
schemes, at considerably smaller computational costs, allow to reduce
unphysical deviations in the total energy up in $100\,000$ times with
respect to those of the standard fourth-order-based iteration approach.

\vspace{6pt}

\noindent
Pacs numbers: 02.60.Cb; 05.10.-a; 95.10.Ce; 95.75.Pq

\end{abstract}

\vspace{13pt}

\begin{multicols}{2}

\section{Introduction}

Understanding the dynamic phenomena in classical and quantum many-body
systems is of importance for the most of areas of physics and chemistry.
The development of efficient algorithms for solving the equations of
motion in such systems should therefore impact a lot of fields of
fundamental research. During the last decade a considerable activity
\cite{Yoshida,Forest,Suzukip,Foresta,Suzukium,Lidis,Dullw,Krech,Omfes}
has been directed on the construction of symplectic time-reversible
algorithms that employ decompositions of the evolution operators
into analytically solvable parts. The decomposition algorithms exactly
preserve all Poincar\'e invariants and, thus, are ideal for long-time
integration in molecular dynamics \cite{Frenkel} and astrophysical
\cite{Wisdom} simulations. The reason is that for these algorithms the
errors in energy conservation appear to be bounded even for relatively
large values of the size of the time step. This is in a sharp contrast
to traditional Runge-Kutta and predictor-corrector schemes \cite{Gear,%
Burden}, where the numerical uncertainties increase linearly with
increasing the integration time \cite{Omfes,Allen,Omelyan,Kinos,Gladm}.

The main attention in previous investigations has been devoted to derive
different-order decomposition algorithms involving only force evaluations
during the time propagation.
For instance, the widely used velocity- and
position-Verlet algorithms \cite{Swope,Tuckerman} relate, in the general
classification, to a three-stages decomposition scheme of the second order
with one force evaluation per step. The fourth-order algorithm by Forest
and Ruth \cite{Forest} corresponds to a scheme with three such force
recalculations and consists of seven single-exponential stages. Sixth-order
schemes are reproduced \cite{Foresta,Lidis} beginning from fifteen stages
and seven evaluations of force for each body in the system per given time
step. With further increasing the order of force decomposition schemes,
the number of stages and thus the number of the corresponding non-linear
equations (which are necessary to solve numerically to obtain the
required time coefficients for single-exponential propagations) increases
drastically. In addition, such equations become too cumbersome and all
these, taking into account the capabilities of modern supercomputers,
led to the impossibility of representing the direct decomposition
algorithms of order eighth and higher in an explicit form \cite{Lidis}.
In order to simplify this problem, it was proposed \cite{Yoshida,Suzukip,%
Suzukium,Lidis,Qin,McLachlan,KahanLi,Murua} to derive higher-order
integrators by composing schemes of lower (actually second) orders.
The resulting second-order-based composition algorithms have been
explicitly obtained up to the tenth order \cite{Suzukium,Lidis,KahanLi}.

Relatively recently \cite{Suzuki,Suzukia,Chin}, a deeper analysis of the
operator factorization process has shown that the class of analytically
integrable decomposition integrators can be extended including additionally
a higher-order commutator into the single-exponential propagations. As a
consequence, a set of new so-called force-gradient algorithms of the
fourth order has been introduced. A distinguishable feature of these
algorithms is the possibility to generate solutions using only positive
values for time coefficients during each substage of the integration.
This is contrary to the original decomposition approach, where beyond
second order (as has been rigorously proved by Suzuki \cite{Suzukip})
any scheme expressed in terms of only force evaluation must produce
some negative time coefficients. We mention that applying negative
time propagations is impossible, in principle, in such important fields
as non-equilibrium statistical mechanics, quantum statistics,
stochastic dynamics, etc., because one cannot simulate diffusion
or stochastic processes backward in time nor sample configurations
with negative temperatures. In the case of stochastic dynamics
simulations it has been demonstrated explicitly \cite{Drozdov,Forbert}
that using fourth-order force-gradient algorithms leads to much
superior propagation over standard Verlet-based schemes of the
second order in that it allows much larger time steps with no
loss of precision. A similar pattern was observed in classical
dynamics simulations comparing the usual fourth-order algorithm
by Forest and Ruth with its force-gradient counterparts \cite{Chin}.

Quite recently, Chin and Kidwell \cite{Chins} has considered a question
of how to iterate the force-gradient algorithms to higher order. The
iteration was based on Creutz's and Gocksch's approach \cite{Creutz}
according to which an algorithm of order $K+2$ can be obtained by
triplet construction of a self-adjoint (i.e. time-reversible) scheme
of order $K$. Then starting from a fourth-order integrator,
it has been shown in
actual celestial mechanics simulations that for orders 6, 8, 10, and
12, the numerical errors corresponding to the force-gradient-based
schemes
are significantly smaller than those of the schemes
basing on iterations of usual non-gradient algorithms.
The resulting efficiency of the integration has
also increased considerably despite an increased computational efforts
spent
on the calculations of force gradients.
The same
has been seen in the case of quantum mechanics simulations when solving
the time-dependent Schr\"odinger equation \cite{ChinChen}.

It is worth emphasizing, however, that the iteration scheme by Chin and
Kidwell is far to be optimal for deriving
high-order
integrators belonging to
the force-gradient class.
The reason is that the number of total force and its gradient
evaluations increases too rapidly with increasing $K$.
Remembering that such evaluations constitute
the most time-consuming part of the calculations, this may restrict
the region of applicability of force-gradient
algorithms to relative low orders only.
Note that high-order computations are especially desirable
in problems of astrophysical interest, because than one can observe
over a system during very long times. They
may also be useful in
highly precise
molecular dynamics and quantum mechanics simulations
to identify or confirm very subtle effects.

In the present paper we propose a general approach to construction of
symplectic force-gradient algorithms of arbitrary orders.
The approach considers the splitting and composing of the
evolution operators on the basic level, taking into account the explicit
structure of truncation terms at each given order in the time step. This
has allowed us to obtain exclusively precise and economical algorithms
with using significantly smaller number of single-exponential propagations
than that appearing within standard decomposition and iteration schemes.
The paper is organized as follows. The
equations of motion
for classical and quantum systems are presented
in section II.A. The integration of these equations by direct
decompositions and their force-gradient generalization are described in
section II.B. Explicit expressions for basic force-gradient algorithms
of orders 2, 4, 6, and 8 are also given there.
The higher-order integration
basing on advanced compositions of lower-order schemes is considered
in section II.C. The composition constants for fourth-, sixth-, and
eighth-order-based schemes are calculated and written down
in the same section up to the overall order 16.
Sections III.A and III.B are devoted to applications of obtained
force-gradient algorithms to molecular dynamics and celestial
mechanics simulations, respectively. A comparative analysis of the new
algorithms with existing integrators is made
there
as well. The final discussion and concluding remarks are
highlighted at the end
in section IV.

\section{General theory of construction of force-gradient algorithms}

\subsection{Basic equations of motion for classical and quantum systems}

Consider first a classical $N$-body system described by the Hamiltonian
\begin{equation}
H = \sum_{i=1}^N \frac{m_i {{\bf v}_i}^2}{2} +
    \frac12 \sum_{i \ne j}^N \varphi(r_{ij}) \equiv T + U \, ,
\end{equation}
where ${\bf r}_i$ is the position of particle $i$ moving with velocity ${\bf
v}_i={\rm d} {\bf r}_i/{\rm d t}$ and carrying mass $m_i$, $\varphi(r_{ij})
\equiv \varphi(|{\bf r}_i-{\bf r}_j|)$ denotes the interparticle potential
of interaction, and $T$ and $U$ relate to the total kinetic and potential
energies, respectively. Then the equations of motion can be presented in
the following compact form
\begin{equation}
\frac{{\rm d} {\mbox{\boldmath $\rho$}}}{{\rm d} t} =
[{\mbox{\boldmath $\rho$}} \circ H] \equiv
L {\mbox{\boldmath $\rho$}}(t) \, .
\end{equation}
Here ${\mbox{\boldmath $\rho$}} = \{ {\bf r}, {\bf v} \} \equiv
\{ {\bf r}_i, {\bf v}_i \}$ is the
full set ($i=1,2,\ldots,N$) of phase variables, $[\,\circ\,]$ represents
the Poisson bracket and
\begin{equation}
L = \sum_{i=1}^N \Big( {\bf v}_i
{\mbox{\boldmath $\cdot$}} \frac{\partial}{\partial {\bf r}_i}+\frac{{\bf
f}_i}{m_i} {\mbox{\boldmath $\cdot$}} \frac{\partial}{\partial {\bf v}_i}
\Big)
\end{equation}
is the Liouville operator with ${\bf f}_i \!=\! -\sum_{j (j \ne i)}^N
\varphi'(r_{ij}) {\bf r}_{ij}/r_{ij}$ being the force acting on particles
due to the interactions.

In the case of quantum systems, the state evolution can be described by
the time-dependent Schr\"odinger equation
\begin{equation}
i \hbar \frac{\partial \psi}{\partial t} = {\cal H} ({\bf r}) \psi
\equiv \Big( {\cal T} + {\cal U}({\bf r}) \Big) \psi \, ,
\end{equation}
where ${\cal T} = -\frac12 \sum_{i=1}^N \hbar^2 {\mbox{\boldmath $\nabla$}}_i^2/
m_i$ and ${\cal U}$ are the kinetic and potential energy operators,
respectively, and $\psi$ is the wave function. So-called
quantum-classical dynamics models \cite{Bornemann} can also be
introduced. This leads to a coupled system of Newtonian (2) and
Schr\"odinger (4) equations. But, in order to simplify notations,
we restrict ourselves to the above purely classic and quantum
considerations.

If an initial configuration ${\mbox{\boldmath $\rho$}}(0)$ or $\psi(0)$ is
provided, the unique solution to Eq.~(2) or (4) can be formally cast as
\begin{equation}
{\mbox{\boldmath $\cal R$}}(t) = {\rm e}^{{\cal L} t} {\mbox{\boldmath
$\cal R$}}(0) \equiv \Big( {\rm e}^{{\cal L} \Delta t} \Big)^l
{\mbox{\boldmath $\cal R$}}(0) \, ,
\end{equation}
where $\Delta t$ and $l=t/\Delta t$ are the size of the single time step
and the total number of steps, respectively, ${\mbox{\boldmath $\cal R$}}$
denotes either ${\mbox{\boldmath $\rho$}}$ or $\psi$, whereas ${\cal L}$
corresponds to $L$ or $-i {\cal H}/\hbar$. As is well known, the time
evolution of many-particle systems
cannot be performed exactly in the general case.
Thus, the problem arises on
evaluating the propagator ${\rm e}^{{\cal L}
\Delta t}$ by numerical methods.

\subsection{Integration by direct decompositions}

\subsubsection{Original decomposition approach}

The main idea of decomposition integration consists in factorization
of the full exponential operator ${\rm e}^{{\cal L} \Delta t}$ on such
subpropagators which allow to be evaluated analytically or at least be
presented in quadratures. Within the original approach, this is achieved
by splitting the operator ${\cal L}={\cal A}+{\cal B}$ into its kinetic
${\cal A}$ and potential ${\cal B}$ parts, where ${\cal A}={\bf v} {\mbox
{\boldmath $\cdot$}} \partial/\partial {\bf r}$ or ${\cal A} = -i {\cal
T}/\hbar$ and ${\cal B}={\bf a} {\mbox{\boldmath $\cdot$}} \partial/
\partial {\bf v}$ with ${\bf a} \equiv \{ {\bf a}_i \} = \{ {\bf f}_i/m_i
\}$ being the
acceleration or ${\cal B}=-i {\cal U}/\hbar$ for the cases of classical
or quantum mechanics, respectively. Then, taking into account the smallness
of $\Delta t$, the total propagator can be decomposed
\cite{Yoshida,Forest,Suzukip,Suzukium,Suzuki} using the formula
\begin{equation}
{\rm e}^{({\cal A}+{\cal B}) \Delta t + {\cal O}(\Delta t^{K+1})} =
\prod_{p=1}^{P} {\rm e}^{{\cal A} a_p \Delta t}
{\rm e}^{{\cal B} b_p \Delta t} \, ,
\end{equation}
where the coefficients $a_p$ and $b_p$ are chosen in such a way to provide
the highest possible value for $K \ge 1$ at a given integer number $P \ge
1$. As a result, integration (5) can performed approximately with the help
of Eq.~(6) by neglecting truncation terms ${\cal O}(\Delta t^{K+1})$. The
precision will increase with increasing the order $K$ and decreasing the
size $\Delta t$ of the time step.

As can be verified readily, the exponential subpropagators ${\rm e}^{{\cal
A} \tau}$ and ${\rm e}^{{\cal B} \tau}$, appearing in the right-hand-side
of Eq.~(6), are analytically integrable for classical systems. Indeed,
taking into account the independence of ${\bf v}$ on ${\bf r}$ and ${\bf
a}$ on ${\bf v}$ yields
\begin{eqnarray}
&&{\rm e}^{{\cal A} \tau} {\mbox{\boldmath $\rho$}} \equiv
{\rm e}^{\tau {\bf v} {\mbox{\boldmath $\cdot$}} \partial/\partial {\bf r}}
\{ {\bf r}, {\bf v} \} = \{ {\bf r} + {\bf v} \tau, {\bf v} \} ,
\nonumber \\ [-5pt] \\ [-5pt]
&&{\rm e}^{{\cal B} \tau} {\mbox{\boldmath $\rho$}} \equiv {\rm e}^{\tau
{\bf a} {\mbox{\boldmath $\cdot$}} \partial/\partial {\bf v}} \{ {\bf r},
{\bf v} \} = \{ {\bf r}, {\bf v} + {\bf a} \tau \} \nonumber
\end{eqnarray}
that represent simple shift operators in position and velocity spaces,
respectively, with $\tau$ being equal to $a_p \Delta t$ or $b_p \Delta t$.
For quantum mechanics
propagations, the kinetic part ${\rm e}^{{\cal A}
\tau} \equiv {\rm e}^{-i \tau {\cal T}/\hbar}$ will require carrying out
two, one direct and one inverse, spatial Fourier transforms \cite{ChinChen},
whereas the calculation of ${\rm e}^{{\cal B} \tau} \equiv {\rm e}^{-i \tau
{\cal U}/\hbar}$ is trivial.

In view of decompositions (6), one can reproduce integrators
of various orders in the time step. In particular, the well-known
second-order ($K=2$) velocity-Verlet algorithm \cite{Tuckerman,Swope}
\begin{equation}
{\rm e}^{({\cal A}+{\cal B}) \Delta t +{\cal O}(\Delta t^3)} =
{\rm e}^{{\cal B} \frac{\Delta t}{2}} {\rm e}^{{\cal A} \Delta t}
{\rm e}^{{\cal B} \frac{\Delta t}{2}} \, ,
\end{equation}
is readily derived from Eq.~(6) by putting $P=2$ with $a_1=0$, $b_1=
b_2=1/2$, and $a_2=1$. The fourth-order ($K=4$) algorithm by Forest and
Ruth \cite{Forest} is obtained from Eq.~(6) at $P=4$ with $a_1=0$,
$a_2=a_4=\theta$, $a_3=(1-2\theta)$, $b_1=b_4=\theta/2$ and $b_2=b_3=
(1-\theta)/2$, where $\theta=1/(2-\sqrt[3]{2})$. Schemes of the sixth
order ($K=6$) are derivable starting from $P=8$
with numerical representation of time
coefficients \cite{Foresta,Lidis}.

The original decomposition approach has, however, a set of disadvantages.
First of all, it is worth pointing out that with further increasing the
order of integration (6) to $K=8$ and higher, the number $2P$ of unknowns
$a_p$ and $b_p$ begins to increase too rapidly. This leads to the
impossibility of representing algorithms of such a type for $K > 6$
in an explicit form \cite{Lidis}, because it becomes impossible to
solve the same number of the resulting cumbersome non-linear equations
(with respect to $a_p$ and $b_p$) even using the capabilities of modern
supercomputers. Another drawback consists in the fact that for $K > 2$
it is impossible \cite{Suzukip} at any $P$ to derive from Eq.~(6) a
decomposition scheme with the help of only positive time coefficients.
For example, in the case of Forest-Ruth integration, three of eight
coefficients,
namely, $a_3$, $b_2$, and $b_3$, are negative. As was mentioned in the
introduction, schemes with negative time coefficients have a restricted
region of application and are not acceptable for simulating non-equilibrium,
quantum statistics, stochastic and other important processes. Moreover, for
schemes expressed in terms of force evaluation only, the main term ${\cal
O}(\Delta t^{K+1})$ of truncation uncertainties appears to be, as a rule,
too big, resulting in decreasing the efficiency of the computations.

\subsubsection{Generalized force-gradient decomposition method}

From the afore said, it is quite desirable to introduce a more general
approach which is free of the above disadvantages. At the same time,
this approach, like the original scheme, must be explicit, i.e., lead
to analytical propagations. In addition, it is expected that the
already known decomposition algorithms should appear from it as
particular cases.

Let us first analyze the structure of third-order truncation errors ${\cal
O}(\Delta t^3)$ of the velocity-Verlet algorithm in detail. Expanding
both the sides of Eq.~(8) into Taylor's series with respect to $\Delta t$,
one finds
\begin{equation}
{\cal O}(\Delta t^3) = \left( \frac{1}{12} [{\cal A},[{\cal A},{\cal B}]] +
\frac{1}{24} [{\cal B},[{\cal A},{\cal B}]] \right) \Delta t^3 +
{\cal O}(\Delta t^5)
\end{equation}
where $[ \ , \ ]$ denotes the commutator of two operators. Taking into
account the explicit expressions for operators ${\cal A}$ and ${\cal B}$
it can be shown that one of the two third-order operators in
Eq.~(9), namely $[{\cal B},[{\cal A},{\cal B}]]$, is
relatively simple and, that is more important, it allows to be handled
explicitly, contrary to the operator $[{\cal A},[{\cal A},{\cal B}]]$.
In the case of classical systems it can be obtained readily
that
\begin{equation}
{\cal C} \equiv [{\cal B},[{\cal A},{\cal B}]] =
\sum_{i=1}^N \frac{{\bf g}_i}{m_i}
{\mbox{\boldmath $\cdot$}} \frac{\partial}{\partial {\bf v}_i}
\equiv {\bf G} {\mbox{\boldmath
$\cdot$}} \frac{\partial}{\partial {\bf v}} \, ,
\end{equation}
where ${\bf g}_{i\alpha} = 2 \sum_{j\beta} {\bf f}_{j\beta}/m_j \partial
{\bf f}_{i\alpha}/\partial {\bf r}_{j\beta}$. In view of the expression
${\bf f}_{i\alpha} = -\sum_{j (j \ne i)} \varphi'(r_{ij}) ({\bf r}_{i
\alpha}-{\bf r}_{j\alpha})/r_{ij}$ for forces, the required force-gradient
evaluations $\partial {\bf f}_{i\alpha}/\partial {\bf r}_{j\beta}$ are
explicitly representable, i.e.,
\begin{eqnarray}
{\bf g}_i = - 2  \sum_{j (j \ne i)}^N
\bigg[
\Big({\bf a}_i-{\bf a}_j \Big) \frac{\varphi'_{ij}}{r_{ij}} +
\frac{{\bf r}_{ij}}{r_{ij}^3} \Big( r_{ij} \varphi''_{ij}-\varphi'_{ij}
\Big)
\nonumber \\ [-9pt] \\ [-9pt]
\times \big({\bf r}_{ij} {\mbox{\boldmath $\cdot$}} ({\bf a}_i-{\bf a}_j)
\big)
\bigg]
\equiv \sum_{j (j \ne i)}^N {\bf g}({\bf r}_{ij})
= {\bf g}_i({\bf r}) \, .
\nonumber
\end{eqnarray}

As can be seen easily from Eqs.~(10) and (11), the operator ${\cal C}$
commutes with ${\cal B} \equiv {\bf a} {\mbox{\boldmath $\cdot$}} \partial/
\partial {\bf v}$,
and, in addition, the function ${\bf G}$ like ${\bf a}$
does not depend on velocity. Then the force-gradient part ${\cal C} \Delta
t^3/24$ of truncation uncertainties (9) can be
extracted by transferring them from the left-hand-side of Eq.~(8) to its
right side and further symmetrically collecting with operator
${\cal B}$
under exponentials. This yields the following force-gradient version
\begin{equation}
{\rm e}^{({\cal A}+{\cal B}) \Delta t +{\cal O}(\Delta t^3)} =
{\rm e}^{{\cal B} \frac{\Delta t}{2} - {\cal C} \frac{\Delta t^3}{48}}
{\rm e}^{{\cal A} \Delta t}
{\rm e}^{{\cal B} \frac{\Delta t}{2} - {\cal C} \frac{\Delta t^3}{48}}
\end{equation}
of the velocity-Verlet integrator, where already
${\cal O}(\Delta t^3) = [{\cal A},[{\cal A},{\cal B}]] \Delta t^3/12$.

In the case of higher-order ($K > 2$) integration (6),
the operator ${\cal C}$ will enter into truncation uncertainties ${\cal O}
(\Delta t^{K+1})$ by various combinations. They can be extracted similarly
as for $K=2$, and we come to a force-gradient
decomposition approach. The most general representation of this
approach is
\begin{equation}
{\rm e}^{({\cal A}+{\cal B}) \Delta t + {\cal O}(\Delta t^{K+1})} =
\prod_{p=1}^{P} {\rm e}^{{\cal A} a_p \Delta t}
{\rm e}^{{\cal B} b_p \Delta t+{\cal C} c_p \Delta t^3} \, ,
\end{equation}
where again at a given $P$ the coefficients $a_p$, $b_p$, as well as
$c_p$ must be chosen in such a way to cancel the truncation terms
${\cal O}(\Delta t^{K+1})$ to the highest possible order $K$. For
$c_p \equiv 0$, generalized factorization (13) reduces to usual
representation (6). It is worth emphasizing that
in view of the velocity independence of ${\bf G}$ on ${\bf v}$,
the modified operator of shifting velocities
remains to be evaluated exactly for any $b_p$ and $c_p$, namely,
\begin{equation}
{\rm e}^{{\cal B} b_p \Delta t+{\cal C} c_p \Delta t^3}
\{ {\bf r}, {\bf v} \} = \{ {\bf r}, {\bf v} + b_p {\bf a} \Delta t
+ c_p {\bf G}  \Delta t^3 \} \, .
\end{equation}
For quantum systems, where ${\cal C} = \sum_i
|{\mbox{\boldmath $\nabla$}}_i {\cal U}|^2$, the corresponding
calculations also present no difficulties (at least for particles
in external fields), because this requires only knowing the gradient
of the potential.

An important feature of decomposition integration (13) is that
it, being applied to classical dynamics simulations, conserves the
symplectic map of particle's flow in phase space. This is so because
separate shifts
of positions (7) and velocities (14) do not change
the phase volume. The time reversibility ${\cal S}(-t) {\mbox
{\boldmath ${\cal R}$}}(t) = {\mbox{\boldmath ${\cal R}$}}(0)$ of
solutions (following from the property ${\cal S}^{-1}(t)={\cal S}(-t)$
of evolution operator ${\cal S}(t)={\rm e}^{{\cal L}t}$) can
be reproduced exactly as well
by imposing additional constraints on the
coefficients $a_p$, $b_p$, and $c_p$. In particular, for velocity-like
decompositions such constraints read: $a_1=0$, $a_{p+1}=a_{P-p+1}$,
$b_p=b_{P-p+1}$, and $c_p=c_{P-p+1}$. Then single-exponential
subpropagators will enter symmetrically into the decompositions,
providing automatically the required reversibility. The case when
the operators of shifting velocity and position are replaced by each
other in the resulting symmetrical decomposition is also possible.
This leads to a position-like integration which can be reproduced
from Eq.~(13) at $a_p=a_{P-p+1}$, $b_p=a_{P-p}$, and $c_p=c_{P-p}$ at
$b_P=0$ and $c_P=0$.

The above symmetry will result in its turn to automatic disappearing
all even-order terms in the error function ${\cal O}(\Delta t^{K+1})$. For
this reason, the order $K$ of time-reversible (self-adjoint) algorithms
may accept only even numbers ($K=2,4,6,\ldots$). The cancellation of
the remaining odd-order terms up to a given order will be provided by
fulfilling a set of basic
conditions for $a_p$, $b_p$, and $c_p$. For example,
the condition $\sum_{p=1}^P a_p=\sum_{p=1}^P b_p=1$ is required to cancel
the first-order truncation uncertainties. Then the error function can be
cast in the form
\begin{eqnarray}
{\cal O}(\Delta t^{K+1}) =
&& {\cal O}_3 \Delta t^3 +
{\cal O}_5 \Delta t^5 +
{\cal O}_7 \Delta t^7 +
\nonumber \\ [-5pt] \\ [-5pt]
&& \ldots +
{\cal O}_{K+1} \Delta t^{K+1} \, .
\nonumber
\end{eqnarray}
In order to kill higher odd-order truncation terms in Eq.~(15),
let us write down explicit expressions for ${\cal O}_3$, ${\cal O}_5$, and
${\cal O}_7$ (this will be enough to derive algorithms
up to the eighth order). Expanding both the sides of Eq.~(13)
into Taylor's series, and collecting the terms with the same powers of
$\Delta t$ one finds:
\begin{equation}
{\cal O}_3 =
\alpha [{\cal A},[{\cal A},{\cal B}]] + \beta [{\cal B},[{\cal A},{\cal B}]] \, ,
\end{equation}
\end{multicols}

\vspace{0pt}

\begin{equation}
{\cal O}_5 =
\gamma_1 [{\cal A},[{\cal A},[{\cal A},[{\cal A},{\cal B}]]]]  +
\gamma_2 [{\cal A},[{\cal A},[{\cal B},[{\cal A},{\cal B}]]]]  +
\gamma_3 [{\cal B},[{\cal A},[{\cal A},[{\cal A},{\cal B}]]]]  +
\gamma_4 [{\cal B},[{\cal B},[{\cal A},[{\cal A},{\cal B}]]]] \, ,
\end{equation}
\begin{eqnarray}
{\cal O}_7 =
\zeta_1 [{\cal B},[{\cal B},[{\cal A},[{\cal B},[{\cal A},[{\cal B},{\cal A}]]]]]]  +
\zeta_2 [{\cal B},[{\cal B},[{\cal B},[{\cal A},[{\cal A},[{\cal B},{\cal A}]]]]]] &+&
\zeta_3 [{\cal B},[{\cal B},[{\cal A},[{\cal A},[{\cal A},[{\cal B},{\cal A}]]]]]]  +
\nonumber \\
\zeta_4 [{\cal B},[{\cal A},[{\cal B},[{\cal A},[{\cal A},[{\cal B},{\cal A}]]]]]]  +
\zeta_5 [{\cal A},[{\cal B},[{\cal B},[{\cal A},[{\cal A},[{\cal B},{\cal A}]]]]]] &+&
\zeta_6 [{\cal A},[{\cal B},[{\cal A},[{\cal B},[{\cal A},[{\cal B},{\cal A}]]]]]]  +
\\
\zeta_7 [{\cal B},[{\cal A},[{\cal A},[{\cal A},[{\cal A},[{\cal B},{\cal A}]]]]]]  +
\zeta_8 [{\cal A},[{\cal B},[{\cal A},[{\cal A},[{\cal A},[{\cal B},{\cal A}]]]]]] &+&
\zeta_9 [{\cal A},[{\cal A},[{\cal B},[{\cal A},[{\cal A},[{\cal B},{\cal A}]]]]]] +
\nonumber \\
\zeta_{10} [{\cal A},[{\cal A},[{\cal A},[{\cal A},[{\cal A},[{\cal B},{\cal A}]]]]]] \, .
\ \ \ \ \ \ \ \ \ \ \ \ \ \ \ \ \ \ \
\ \ \ \ \ \ \ \ \ \ \ \ \ \ \ \ \ \ \
\nonumber
\end{eqnarray}
\begin{multicols}{2}
\noindent
Here we take into account the fact
that the operators ${\cal B}$ and ${\cal C}$ commute
between themselves, i.e. $[{\cal B}, {\cal C}]=0$, so that any occurrence of
constructions containing the chain $[{\cal B},[{\cal B},[{\cal A},{\cal B}]]]$
has been ignored
(in particular for fifth-order truncation term ${\cal O}_5$
this has allowed us to exclude the two zero-valued commutators $[{\cal B},
[{\cal B},[{\cal B},[{\cal A},{\cal B}]]]]$ and $[{\cal A},[{\cal B},[{\cal
B},[{\cal A},{\cal B}]]]]$).
The multipliers $\alpha$, $\beta$, $\gamma_{1-4}$, and $\zeta_{1-10}$,
arising in Eq.~(16)--(18), are functions of the coefficients $a_p$, $b_p$,
and $c_p$, where $p=1,2,\ldots,P$. The concrete form of these functions
will depend on $P$ and the version (velocity or position) under
consideration.

The most simple way to obtain explicit expressions for
the multipliers consists in the following. First, since we are dealing
with self-adjoint schemes, the total number of single-exponential
operators (stages) in Eq.~(13) is actually equal to $S=2P-1$, i.e.
it
accepts only odd values (mention that one of the boundary set of
coefficients is set to zero, $a_1=0$ or $b_P=c_P=0$). Then we can always
choose a central single-exponential operator, and further consecutively
applying $P-1$ times the two types of symmetric transformation
$$
{\rm e}^{{\cal W}^{(n+1)} + {\cal O}^{(n+1)}} =
{\rm e}^{{\cal A} a^{(n)} \Delta t}
{\rm e}^{{\cal W}^{(n)} + {\cal O}^{(n)}}
{\rm e}^{{\cal A} a^{(n)} \Delta t}
$$
\vspace{-30pt}
\begin{equation}
\end{equation}
\vspace{-30pt}
\begin{eqnarray*}
&& {\rm e}^{{\cal W}^{(n+1)} + {\cal O}^{(n+1)}} =
\\ [4pt]
&& {\rm e}^{{\cal B} b^{(n)} \Delta t+{\cal C} c^{(n)} \Delta t^3}
{\rm e}^{{\cal W}^{(n)} + {\cal O}^{(n)}}
{\rm e}^{{\cal B} b^{(n)} \ \Delta t+{\cal C} c^{(n)} \Delta t^3}
\end{eqnarray*}
come to factorization (13), where
$$
{\cal W} = (\nu {\cal A} + \sigma {\cal B}) \Delta t
$$
and ${\cal O}$ is defined by Eq.~(15). The quantities $a^{(n)}$, $b^{(n)}$,
and $c^{(n)}$ are related to $a_p$, $b_p$, and $c_p$, respectively (the
relationship between $n$ and $p$ is determined below). For velocity-like
decomposition with even $P$ or position-like at odd $P$, the central
operator is correspondingly ${\rm e}^{{\cal A} a_{(P-2)/2+1} \Delta t}$
or ${\rm e}^{{\cal A} a_{(P-1)/2+1} \Delta t}$. So that here we must put
$\sigma^{(0)}=0$ as well as $\alpha^{(0)}=\beta^{(0)}=\gamma_{1-4}^{(0)}=
\zeta_{1-10}^{(0)}=0$ and let either $\nu^{(0)}=a_{(P-2)/2+1}$ or
$\nu^{(0)}=a_{(P-1)/2+1}$ on the very beginning ($n=0$) of the recursive
procedure. The start of the procedure should be performed with the second
line of Eq.~(19) at $b^{(0)}=b_{(P-2)/2}$ and $c^{(0)}=c_{(P-2)/2}$ or
$b^{(0)}=b_{(P-1)/2}$ and $c^{(0)}=c_{(1-2)/2}$ with further decreasing
the index $p$ with increasing the number $n=1,2,\ldots P-1$
at $a^{(n)} \equiv a_p$,
$b^{(n)} \equiv b_p$, and $c^{(n)} \equiv c_p$ in both the lines of
transformation (19). For velocity-like decomposition with odd $P$ or
position-like at even $P$, the central operator will be ${\rm e}^{{\cal
B} b_{(P-1)/2+1} \Delta t+{\cal C} c_{(P-1)/2+1} \Delta t^3}$ or ${\rm
e}^{{\cal B} b_{(P-2)/2+1} \Delta t+{\cal C} c_{(P-2)/2+1} \Delta t^3}$,
corresponding to $\sigma^{(0)}=b_{(P-1)/2+1}$ and $\beta^{(0)}=c_{(P-1)/2
+1}$ or $\sigma^{(0)}=b_{(P-2)/2+1}$ and $\beta^{(0)}=c_{(P-2)/2+1}$,
respectively,
with
$\nu^{(0)}=0$ and $\alpha^{(0)}=\gamma_{1-4}^{(0)}=\zeta_{1-10}^{(0)}=0$.
In this case, the procedure
should be started with the first type of
transformation
at $a^{(0)}=b_{(P-1)/2+1}$ or $a^{(0)}=b_{(P-2)/2+1}$
with decreasing $p$ at increasing $n$ for $b^{(n)} \equiv b_p$, $c^{(n)}
\equiv c_p$, and $a^{(n)} \equiv a_p$ in Eq.~(19).

The recursive relations between the multipliers $\nu$, $\sigma$,
$\alpha$, $\beta$, and $\gamma_{1-4}$ corresponding to the first
line of Eq.~(19) are:
\begin{equation}
\nu^{(n+1)}    =\nu^{(n)}    + 2 a^{(n)} \, ,
\ \ \ \ \ \
\sigma^{(n+1)} =\sigma^{(n)} \, ,
\end{equation}
\begin{equation}
\alpha^{(n+1)} =\alpha^{(n)} - a^{(n)} \sigma^{(n)} \big(a^{(n)} +
\nu^{(n)}\big)/6
\, ,
\end{equation}
\begin{equation}
\beta^{(n+1)}  =\beta^{(n)} -     a^{(n)} {\sigma^{(n)}}^2/6 \, ,
$$
\end{equation}
\end{multicols}
$$
\gamma_1^{(n+1)}=\gamma_1^{(n)}+
     a^{(n)} \big(a^{(n)} + \nu^{(n)}\big) \big(
\big(7 ({a^{(n)}}^2 + 7 a^{(n)} \nu^{(n)} +
{\nu^{(n)}}^2\big) \sigma^{(n)}
-60 \alpha^{(n)}
\big)/360
\, ,
$$
$$
\gamma_2^{(n+1)}=\gamma_2^{(n)}+
     a^{(n)} \big(
30 \alpha^{(n)} \sigma^{(n)}
-30 a^{(n)} \beta^{(n)} - 30 \beta^{(n)} \nu^{(n)}
+ 3 ({a^{(n)}}^2 {\sigma^{(n)}}^2 +
   2 a^{(n)} \nu^{(n)} {\sigma^{(n)}}^2 +
{\nu^{(n)}}^2 {\sigma^{(n)}}^2\big)/180
\, ,
$$
\vspace{-18pt}
\begin{equation}
\end{equation}
\vspace{-23pt}
$$
\gamma_3^{(n+1)}=\gamma_3^{(n)}+
     a^{(n)} \sigma^{(n)} \big(
\big(8 ({a^{(n)}}^2 + 12 a^{(n)} \nu^{(n)} +
{\nu^{(n)}}^2\big) \sigma^{(n)}
-120 \alpha^{(n)}
\big)/360
\, ,
$$
$$
\gamma_4^{(n+1)}=\gamma_4^{(n)}+
     a^{(n)} \sigma^{(n)} \big(
\big(6 a^{(n)} + \nu^{(n)}\big) {\sigma^{(n)}}^2
-60 \beta^{(n)}
\big)/180
\, .
$$
\begin{multicols}{2}
\noindent
For the second type of transformation the relations read:
\begin{equation}
\nu^{(n+1)}    =\nu^{(n)} \, ,
\ \ \ \ \ \
\sigma^{(n+1)} =\sigma^{(n)} + 2 b^{(n)}
\, ,
\end{equation}
\begin{equation}
\alpha^{(n+1)} =\alpha^{(n)} +      b^{(n)} {\nu^{(n)}}^2/6
\, ,
\end{equation}
\begin{equation}
\beta^{(n+1)}  =\beta^{(n)} + \big(12 c^{(n)} +
b^{(n)} \nu^{(n)} \big(b^{(n)} + \sigma^{(n)}\big)\big)/6
\, ,
\end{equation}
\end{multicols}

\vspace{-9pt}

$$
\gamma_1^{(n+1)}=\gamma_1^{(n)} -      b^{(n)} {\nu^{(n)}}^4/360
\, ,
$$
$$
\gamma_2^{(n+1)}=\gamma_2^{(n)}
-     \nu^{(n)} \big(60 \alpha^{(n)} b^{(n)} -
\nu^{(n)} \big( 30 c^{(n)} - b^{(n)} \nu^{(n)} \big(6 b^{(n)} +
\sigma^{(n)}\big)\big)\big)/180
\, ,
$$
\begin{equation}
\gamma_3^{(n+1)}=\gamma_3^{(n)}+
     b^{(n)} \nu^{(n)} \big(60 \alpha^{(n)} +
{\nu^{(n)}}^2 \big(4 b^{(n)} - \sigma^{(n)}\big)\big)/360
\, ,
\end{equation}
\begin{eqnarray*}
\gamma_4^{(n+1)}=\gamma_4^{(n)}-
&&
\big(
 30 \alpha^{(n)} b^{(n)} \big(b^{(n)} + \sigma^{(n)}\big) -
\nu^{(n)} \big(30 \beta^{(n)} b^{(n)} + 60 b^{(n)} c^{(n)}
-
\\ [3pt]
&&
3 {b^{(n)}}^3 \nu^{(n)} +
    30 c^{(n)} \sigma^{(n)} - 2 {b^{(n)}}^2 \nu^{(n)} \sigma^{(n)} -
b^{(n)} \nu^{(n)} {\sigma^{(n)}}^2\big)\big)/180
\, .
\end{eqnarray*}
\begin{multicols}{2}
\noindent
The relations for $\zeta_{1-10}$ are presented in Appendix. In such a
way, at the end of the recursive process (i.e. after $P-1$ steps)
the multipliers can readily be obtained. The form of the first two
multipliers are particularly simple and look as $\nu=\sum_{p=1}^P a_p$
and $\sigma=
\sum_{p=1}^P b_p$. So that, as was already mentioned above, putting
$\nu=1$ and $\sigma=1$ will cancel the first-order truncation
uncertainties (because the resulting exponential propagator must
behave like ${\rm e}^{({\cal A}+{\cal B}) \Delta t}$). Next multipliers
should be set to zero and we come to the necessity of solving a system
of non-linear equations (so-called order conditions) with respect to
$a_p$, $b_p$, and $c_p$. We shall now consider actual self-adjoint
algorithms of orders $K=2$, 4, 6, and 8.

\subsubsection{Force-gradient algorithms of order two}

Putting $P=2$ in Eq.~(13) with $a_1=0$, $b_1=b_2=1/2$, $a_2=1$, and
$c_1=c_2 \equiv \xi$
leads to the following velocity-force-gradient
algorithm of the second ($K=2$) order,
\begin{equation}
{\rm e}^{({\cal A}+{\cal B}) \Delta t + {\cal O}_3 \Delta t^3} =
{\rm e}^{{\cal B} \frac{\Delta t}{2} + \xi {\cal C} \Delta t^3}
{\rm e}^{{\cal A} \Delta t}
{\rm e}^{{\cal B} \frac{\Delta t}{2} + \xi {\cal C} \Delta t^3} \, ,
\end{equation}
with $\alpha=1/12$, and $\beta=1/24 + 2 \xi$. Note that
here and below, for reducing the number of unknowns,
we will always take into account in advance the symmetry of coefficients
$a_p$, $b_p$, and $c_p$ as well as the fulfilling the first-order
conditions $\nu=\sum_{p=1} a_p=1=\sum_{p=1} b_p=1=\sigma$ when writing
decomposition formulas.
Then solving the equation
$\beta=0$ yields $\xi=-1/48$ and we come to the already
found integrator (12). It is worth remarking
that negative values of quantities $c_p$
at force gradients have nothing to do with the above problem of
positiveness of time coefficients arising at velocities and
forces, i.e., for $a_p$ and $b_p$. The reason is that the incremental
velocity $b_p {\bf a} \Delta t + c_p {\bf G} \Delta t^3$ in Eq.~(14)
can be rewritten as $(b_p {\bf a} + c_p {\bf G} \Delta t^2) \Delta t
\equiv b_p {\bf \tilde a} \Delta t$, and thus treated as the velocity
changing in a modified step-size-dependent acceleration field
${\bf \tilde a}={\bf a}+\frac{c_p}{b_p} {\bf G}  \Delta t^2$.

The position counterpart of Eq.~(28) is obtained from Eq.~(13) at $a_1=0$,
$a_1=a_2=1/2$, $b_1=1$, $b_2=0$, $c_1 \equiv \xi$ and $c_2=0$, that yields
\begin{equation}
{\rm e}^{({\cal A}+{\cal B}) \Delta t + {\cal O}_3 \Delta t^3} =
{\rm e}^{{\cal A} \frac{\Delta t}{2}}
{\rm e}^{{\cal B} \Delta t + \xi {\cal C} \Delta t^3}
{\rm e}^{{\cal A} \frac{\Delta t}{2}}
\end{equation}
for which $\alpha=-1/24$ and $\beta=-1/12 + \xi$.
Letting $\xi=1/12$ will minimize the third-order truncation errors to the
value $\alpha [{\cal A},[{\cal A}, {\cal B}]] \Delta t^3$ which is even
twice smaller in magnitude than that of the velocity version. Note, however,
that for both versions (28) and (29), which require one force plus one
force-gradient evaluations per time step,
the order of integration is not
increased with respect to the usual (when $\xi=0$)
Verlet integrators
requiring only one force recalculation.
In view of this,
the applying force gradients
in a particular case of $P=2$ can be justified
only for strongly interacting systems when the kinetic part ${\cal A}$
of the Liouville operator ${\cal L}$ is much smaller than the potential
part ${\cal B}$, i.e., when ${\cal L}=\varepsilon {\cal A} + {\cal B}$ with
$\varepsilon \ll 1$. Then the remaining part $\alpha [{\cal A},[{\cal A},
{\cal B}]] \Delta t^3$ of local uncertainties will behave like $\propto
\varepsilon^2$ and can be neglected.

\subsubsection{Force-gradient algorithms of order four}

Further
increasing $P$ on unity allows us to kill exactly
both the multipliers $\alpha$ and
$\beta$, that is needed for obtaining fourth-order ($K=4$) integrators. So
that choosing $P=3$ leads to the velocity-like propagation
\begin{eqnarray}
&& {\rm e}^{({\cal A}+{\cal B}) \Delta t +{\cal O}(\Delta t^5)} =
\nonumber \\ [-12pt] \\ [2pt]
&& {\rm e}^{{\cal B} \lambda \Delta t + \xi {\cal C} \Delta t^3}
{\rm e}^{{\cal A} \frac{\Delta t}{2}}
{\rm e}^{{\cal B} (1 - 2 \lambda) \Delta t + \chi {\cal C} \Delta t^3}
{\rm e}^{{\cal A} \frac{\Delta t}{2}}
{\rm e}^{{\cal B} \lambda \Delta t + \xi {\cal C} \Delta t^3}
\ \ \ \
\nonumber
\end{eqnarray}
following from Eq.~(13) at $a_1=0$, $a_2=a_3=1/2$, $b_1=b_3=\lambda$,
$b_2=1 - 2 \lambda$, $c_1=c_3=\xi$, and $c_2=\chi$. Here relations (21),
(22), (25), and (26) come to the two order conditions
$$
\alpha=-\frac{1 - 6 \lambda}{24} = 0
\, , \ \ \
\beta=-\frac{1}{12} + \frac{\lambda}{2} - \frac{\lambda^2}{2} +
2 \xi + \chi = 0 \, ,
$$
with three unknowns $\lambda$, $\xi$, and $\chi$. The first unknown is
immediately obtained satisfying the first condition,
\begin{equation}
\lambda=\frac16 \, .
\end{equation}
The second equality is then reduces to $2\xi+\chi=1/72$, resulting in a
whole family of velocity-force-gradient algorithms of the fourth order.
In general, such algorithms will require two force and two force-gradient
recalculations per time step.

Remembering that we are interested
in the derivation of most efficient integrators, three cases deserve
to be considered. The two of them are aimed to reduce the number
of force-gradient recalculations from two to one. This is possible by
choosing either
\begin{equation}
\xi=0 \, , \ \ \ \ \ \ \ \chi=\frac{1}{72} \ \
\end{equation}
or
\begin{equation}
\chi=0 \, , \ \ \ \ \ \ \ \xi=\frac{1}{144} \, .
\end{equation}
In the third case we will try to minimize the norm
\begin{equation}
\gamma=\sqrt{\gamma_1^2+\gamma_2^2+\gamma_3^2+\gamma_4^2}
\end{equation}
of fifth-order truncation errors ${\cal O}(\Delta t^5)$ at $\xi \ne 0$ and
$\chi = 1/72 - 2 \xi \ne 0$, treating $\xi$ as a free parameter. In view of
recursive relations (23) and (27), explicit expressions for the components
of ${\cal O}(\Delta t^5)$ are
$$
\gamma_1\!=\!\frac{7 - 30 \lambda}{5760}
, \ \
\gamma_2\!=\!\frac{1}{480}-\frac{\chi}{24}-\frac{\lambda^2}{24}+\frac{\xi}{6}
, \ \
\gamma_3\!=\!\frac{1}{360} - \frac{\lambda}{48} + \frac{\lambda^2}{24}
,
$$
$$
\gamma_4=\frac{1}{120} - \frac{\lambda}{16} + \frac{7 \lambda^2}{48} -
\frac{\lambda^3}{8} + \frac{\xi}{6} - \frac{\chi}{2} \bigg(\frac{1}{3}
- \lambda \bigg) \, .
$$
Then taking into account Eq.~(31) one finds the function
$$
\gamma=\frac{1}{135 \sqrt{2048}} \sqrt{19 + 12240 \xi + 6480000 \xi^2}
$$
with the minimum $\gamma_{\rm min}= \sqrt{661}/43200 \approx 0.000595$ at
\begin{equation}
\xi = -\frac{17}{18000} \, , \ \ \ \ \ \
\chi = \frac{71}{4500} \, .
\end{equation}
At the same time, the values of $\gamma$ corresponding to first two
algorithms (32) and (33) constitute
$ \sqrt{19/2048}/135 \approx 0.000713$
and $7 \sqrt{17}/8640 \approx 0.00334$,
respectively.

Position version of (30) reads
\begin{eqnarray}
&& {\rm e}^{({\cal A}+{\cal B}) \Delta t +{\cal O}(\Delta t^5)} =
\nonumber \\ [-5pt] \\ [-5pt]
&& {\rm e}^{{\cal A} \lambda \Delta t}
{\rm e}^{{\cal B} \frac{\Delta t}{2} + \xi {\cal C} \Delta t^3}
{\rm e}^{{\cal A} (1 - 2 \lambda) \Delta t}
{\rm e}^{{\cal B} \frac{\Delta t}{2} + \xi {\cal C} \Delta t^3}
{\rm e}^{{\cal A} \lambda \Delta t}
\nonumber
\ \ \ \
\end{eqnarray}
and
is obtained from Eq.~(13) at $P=3$ with
$a_1=a_3=\lambda$, $a_2=(1 - 2 \lambda)$, $b_1=b_2=1/2$, $c_1=c_2=\xi$,
and $b_3=c_3=0$. Here, the number of unknowns coincides with the number
of the order conditions
$$
\alpha=\frac{1}{12} - \frac{\lambda}{2} + \frac{\lambda^2}{2} = 0
\, , \ \ \ \ \ \
\beta=\frac{1}{24} - \frac{\lambda}{4} + 2 \xi = 0
$$
solving of which yields two solutions,
\begin{equation}
\lambda = \frac12 \left( 1 \mp \frac{1}{\sqrt{3}} \right)
\, , \ \ \ \ \ \
\xi = \frac{1}{48} \left( 2 \mp \sqrt{3} \right)
\, .
\end{equation}
Then the norm of truncation uncertainties
${\cal O}(\Delta t^5)$ appearing in Eq.~(36)
is
$
\gamma=(1873 \mp 40 \sqrt{2187})^{1/2}/2160
$,
so that
the preference should be given to sign ``--'' in Eq.~(37), because
this leads to a smaller value, $\gamma_- \approx 0.000715$, of
$\gamma$ (whereas $\gamma_+ \approx 0.0283$). Position algorithm (36)
needs, like velocity version (35), in two force and the same
number of force-gradient evaluations per time step.

Integrators (32) and (37) have been previously derived by Suzuki \cite{Suzuki}
based on McLachlan's method of small perturbation \cite{McLachlana}
and referred by Chin \cite{Chin} to schemes A and B, respectively.
Algorithms (33) and (35)
are new and will labeled by us as A$'$ and A$''$. While
scheme A$'$ seems has no advantages over the A-integrator,
the new algorithm A$''$
corresponds to the best accuracy
of the integration,
because it minimizes $\gamma$. It requires, however, one
extra force-gradient evaluation and, thus, can be recommended for
situations when this evaluation does not present significant
difficulties.

With the aim
of considerable decreasing the truncation errors in a little
additional computational efforts, Chin \cite{Chin} has proposed to
consider extended force-gradient algorithms of the fourth order. This
has been achieved by increasing the number of force recalculations on
unity with respect to the necessary minimum, i.e. choosing $n_{\rm f}=
3$. At the same time, the number of force-gradient evaluations was
fixed to its minimal value $n_{\rm g}=1$. Within our general approach,
it is possible to introduce two fourth-order
schemes satisfying the above requirements. The schemes are
\begin{eqnarray}
{\rm e}^{({\cal A}+{\cal B}) \Delta t +{\cal O}(\Delta t^5)} =
{\rm e}^{{\cal A} \theta \Delta t}
{\rm e}^{{\cal B} \lambda \Delta t}
{\rm e}^{{\cal A} (1 - 2 \theta) \frac{\Delta t}{2}} &&
\nonumber \\ [-6pt] \\ [-6pt]
\times
{\rm e}^{{\cal B} (1 - 2 \lambda) \Delta t + \chi {\cal C} \Delta t^3}
{\rm e}^{{\cal A} (1 - 2 \theta) \frac{\Delta t}{2}}
{\rm e}^{{\cal B} \lambda \Delta t}
{\rm e}^{{\cal A} \theta \Delta t} &&
\nonumber
\end{eqnarray}
and
\begin{eqnarray}
{\rm e}^{({\cal A}+{\cal B}) \Delta t +{\cal O}(\Delta t^5)} =
{\rm e}^{{\cal B} \lambda \Delta t + \xi {\cal C} \Delta t^3}
{\rm e}^{{\cal A} \theta \Delta t}
{\rm e}^{{\cal B} (1 - 2 \lambda) \frac{\Delta t}{2}}
\nonumber \\ [-6pt] \\ [-6pt]
\times
{\rm e}^{{\cal A} (1 - 2 \theta) \Delta t}
{\rm e}^{{\cal B} (1 - 2 \lambda) \frac{\Delta t}{2}}
{\rm e}^{{\cal A} \theta \Delta t}
{\rm e}^{{\cal B} \lambda \Delta t + \xi {\cal C} \Delta t^3}
\nonumber
\end{eqnarray}
following from Eq.~(13) at $P=4$ and corresponding to position- and
velocity-like integration, respectively. Note that
further we will not present
the relationship between the coefficients $a_p$, $b_p$, $c_p$ of
Eq.~(13) and reduced variables (such as, for example, $\theta$, $\lambda$,
$\chi$ in Eq.~(38)) in view of its evidence.

The order conditions for scheme (38) are
\begin{eqnarray*}
\alpha&=&-\frac{1}{24} + \lambda
\bigg( \frac{1}{4} - \theta + \theta^2 \bigg) = 0 \, ,
\\ [-2pt] \\ [-2pt]
\beta&=&-\frac{1}{12} + \frac{\lambda}{2} - \frac{\lambda^2}{2}
- \lambda \theta \Big( 1 - \lambda \Big) + \chi = 0
\end{eqnarray*}
and solving them one obtains
\begin{equation}
\theta = \frac{1}{2} \pm \frac{1}{\sqrt{24 \lambda}}
\, , \ \ \ \ \ \
\chi = \frac{ 1 \pm \sqrt{6 \lambda} (1 - \lambda)}{12}
\, .
\end{equation}
Relations (40) constitute a family of extended force-gradient
position algorithms (38) of the fourth order with $\lambda$ being a
free parameter. Chin \cite{Chin} has introduced an algorithm like (38)
in somewhat another way, namely, as a symmetric product of two third-order
schemes. This results only in one set of time coefficients which
can be reproduced (at sign ``--'') from Eq.~(40) as a particular case
corresponding to
\begin{equation}
\lambda=\frac{3}{8} \, , \ \ \ \ \ \
\theta=\frac{1}{6} \, , \ \ \ \ \ \
\chi=\frac{1}{192}
\end{equation}
and has been referred to scheme C.

Solution (41) may not, however, be necessarily optimal in view of the fact
that it
does not minimize the norm $\gamma$ (see Eq.~(34)) of truncation uncertainties
${\cal O}(\Delta t^5)$. Indeed, the components of $\gamma$ for scheme
(38) are
\begin{eqnarray*}
\gamma_1&=&-\frac{1}{1920} + \frac{1}{6912 \lambda}
\, , \ \ \
\gamma_2=\frac{6 \pm 5 \sqrt{6 \lambda}}{2880}
\, ,
\\ [2pt]
\gamma_3&=&-\frac{1}{360} \bigg( \frac{3}{2} \pm
\frac{5}{\sqrt{96 \lambda}}
\pm 5 \sqrt{\frac{\lambda}{24}}
\bigg)
\, ,
\\ [2pt]
\gamma_4&=&-\frac{1}{1440} \Big( 3 \pm
5 \sqrt{24 \lambda} + 45 \lambda - 30 \lambda^2
\Big)
\, ,
\end{eqnarray*}
where Eq.~(40) has been used to express the function
$\gamma(\lambda)$ in terms of
one parameter $\lambda$ exclusively. The global minimum
of this function is $\gamma_{\rm min} \approx 0.000141$
and achieved (at sign minus) at
\begin{eqnarray}
\lambda &=&  0.2470939580390842{\rm E}\!+\!00
\, , \ {\rm and \ thus}
\nonumber
\\
\theta  &=&  0.8935804763220157{\rm E}\!-\!01
\, ,
\\
\chi    &=&  0.6938106540706989{\rm E}\!-\!02
\nonumber
\end{eqnarray}
(all results found numerically will be presented within sixteen
significant digits for schemes up to the eighth order and
within thirty two digits for order ten and higher).
On the other hand,
the value of $\gamma$ corresponding to scheme C (Eq.~(41)),
is equal only to $\sqrt{87817}/414720 \approx 0.000715$, i.e. it
is approximately in 5 times larger than that of the optimized
algorithm (42). The last algorithm we will designate
as scheme C$'$.

A similar pattern is observed in the case of extended
velocity-force-gradient integration (39). Previously Chin and Chen
\cite{ChinChen} have indicated that for quantum mechanics simulations
the integration of such a type is more preferable than position-like
scheme (38), because it requires a fewer number of spatial Fourier
transforms. Again using the symmetric product of two third-order
integrators to increase the order from three to four, they have
obtained the following set
\begin{equation}
\lambda = \frac{1}{8} \, , \ \ \ \ \ \
\theta  = \frac{1}{3} \, , \ \ \ \ \ \
\xi     = \frac{1}{384}
\end{equation}
of time coefficients and referred it to scheme D. We have realized that
this set is not only possible and found a whole family of solutions
(which includes (43)),
namely,
$$
\lambda=  \frac{1}{12} \bigg( 6 + \frac{1}{\theta (\theta-1)} \bigg)
\, , \ \ \ \ \ \
\xi = -\frac{1}{288} \bigg(  6
- \frac{1}{\theta (\theta-1)^2} \bigg)
\, ,
$$
where $\theta$ should be considered as a free parameter. The optimal
solution, which minimizes the norm $\gamma$
of fifth-order errors to the value
$\gamma_{\rm min} \approx 0.000855$, is
\begin{eqnarray}
\lambda &=&  0.4432204907934768{\rm E}\!-\!01
\, , \nonumber \\
\theta  &=&  0.2409202729169543{\rm E}\!+\!00
\, , \\
\xi     &=&  0.4179297897540420{\rm E}\!-\!02
\nonumber
\end{eqnarray}
and will be labeled as scheme D$'$.
At the same time, the norm of errors corresponding to
scheme D (Eq.~(43)) is equal to $\gamma =\sqrt{237457}/414720 \approx
0.00117$, i.e. it exceeds the minimum, that may results in
decreasing the precision of the calculations.

As can be ensured readily, the time coefficients arising at basic
operators ${\cal A}$ and ${\cal B}$ under exponentials are positive
for all the fourth-order force-gradient algorithms described in this
subsection. Therefore, contrary to usual force Forest-Ruth-like schemes,
such
algorithms can simulate dynamical processes in all areas of physics
and chemistry without any principal restrictions.

\subsubsection{Force-gradient algorithms of order six}

Beginning from $P=5$, the force-gradient factorization
being written in velocity representation allows to eliminate the components
of truncation uncertainties up to the sixth order ($K=6$)
inclusively. In view of Eq.~(13), such a representation reads
\begin{eqnarray}
{\rm e}^{({\cal A}+{\cal B}) \Delta t +{\cal O}(\Delta t^7)} =
{\rm e}^{{\cal B} \vartheta \Delta t + \mu {\cal C} \Delta t^3}
{\rm e}^{{\cal A} \theta \Delta t}
{\rm e}^{{\cal B} \lambda \Delta t + \xi {\cal C} \Delta t^3}
&&
\nonumber \\
\times
{\rm e}^{{\cal A} (1 - 2 \theta) \frac{\Delta t}{2}}
{\rm e}^{{\cal B} \big(1 - 2 (\lambda+\vartheta) \big) \Delta t +
\chi {\cal C} \Delta t^3}
{\rm e}^{{\cal A} (1 - 2 \theta) \frac{\Delta t}{2}}
&&
\\
\times
{\rm e}^{{\cal B} \lambda \Delta t + \xi {\cal C} \Delta t^3}
{\rm e}^{{\cal A} \theta \Delta t}
{\rm e}^{{\cal B} \vartheta \Delta t + \mu {\cal C} \Delta t^3}
&& .
\nonumber
\end{eqnarray}
The number of unknowns in propagation (45) is the same as
the number of order conditions which now take the form
\end{multicols}
$$
\alpha = \lambda \bigg( \frac{1}{4} - \theta + \theta^2 \bigg)
- \frac{1}{24} \bigg( 1 - 6 \vartheta \bigg) = 0 \, ,
$$
\begin{eqnarray*}
\beta =
\chi - \frac{1}{12} \bigg( 1 - 24 \mu - 6 \lambda^2 (2 \theta - 1)
- 6 \vartheta +
6 \vartheta^2 -
   6 \lambda (2 \theta - 1) (2 \vartheta - 1) - 24 \xi \bigg) = 0 \, ,
\end{eqnarray*}
\begin{eqnarray*}
\gamma_1 = \frac{1}{5760} \bigg( 7 - 30 \lambda (2 \theta - 1)^2
( 1 + 4 \theta - 4 \theta^2) - 30 \vartheta \bigg) = 0 \, ,
\end{eqnarray*}
\begin{eqnarray}
\gamma_2 = && \frac{1}{480}
\bigg( 1 - 20 \chi + 80 \mu - 20 \lambda^2 ( 1
- 8 \theta + 18 \theta^2 - 12 \theta^3) -
  20 \vartheta^2 +
\nonumber \\ [-8pt] \\ [-8pt]
&&
20 \lambda
(2 \theta -1) ( \theta + 2 \vartheta - 6 \theta \vartheta) +
80 \xi
- 480 \theta \xi + 480 \theta^2 \xi \bigg) = 0 \, ,
\nonumber
\end{eqnarray}

\vspace{-6pt}

\begin{eqnarray*}
\gamma_3 = \frac{1}{720}
\bigg( 2 - 30 \lambda^2 (2 \theta - 1)^3
- 15 \vartheta + 30 \vartheta^2 -
  15 \lambda (2 \theta - 1)^2 ( 1
- 4 \vartheta - \theta (4 \vartheta - 2) ) \bigg) = 0 \, ,
\end{eqnarray*}

\vspace{-6pt}

\begin{eqnarray*}
\gamma_4 = &&
\frac{1}{240}
\bigg( 2 + 40 \mu - 30 \lambda^3 (2 \theta - 1)^2 -
  40 \chi (1 + \lambda (6 \theta - 3)
- 3 \vartheta) -
15 \vartheta +
  35 \vartheta^2 - 30 \vartheta^3 -
5 \lambda^2 (2 \theta - 1)
\\
&&
\times
   (7 -
18 \vartheta +
6 \theta (2 \vartheta - 1)) +
  5 \lambda (2 \theta - 1) (3
- 14 \vartheta + 18 \vartheta^2 +
    2 \theta (1 - 6 \vartheta
+ 6 \vartheta^2)) + 40 \xi - 240 \theta \xi +
  480 \theta \vartheta \xi \bigg) = 0 \, .
\end{eqnarray*}
\begin{multicols}{2}
The unique real solution to system (46) is
$$
\theta = \frac{1}{2}+\frac{\sqrt[3]{675+75 \sqrt{6}}}{30}
+\frac{5}{2 \sqrt[3]{675+75 \sqrt{6}}}
\, , \ \ \
\vartheta = \frac{\theta}{3}
\, ,
$$
\begin{equation}
\lambda = - \frac{5 \theta}{3} \bigg( \theta - 1 \bigg)
\, , \ \ \ \ \ \
\xi = - \frac{5 \theta^2}{144} + \frac{\theta}{36} - \frac{1}{288}
\, ,
\ \ \
\end{equation}
$$
\chi = \frac{1}{144} - \frac{\theta}{36} \bigg( \frac{\theta}{2} + 1
\bigg)
\, , \ \ \ \ \ \ \ \ \
\mu = 0 \, .
\ \ \ \ \
$$
Solution (47) constitutes a velocity-force-gradient algorithm of the
sixth order with four force and three (since $\mu=0$) force-gradient
evaluations per time step, i.e., with $n_{\rm f}=4$ and $m_{\rm g}=3$.
Its advantage over usual sixth-order integrators consists in the fact
that it is composed from a considerably smaller number, namely $S=2P-1
=9$, instead of 15, of single exponential operators. The norm
\begin{equation}
\zeta=
\sqrt{\sum_{k=1}^{10} \zeta_k^2}
\end{equation}
of seventh-order truncation errors ${\cal O}(\Delta t^7)$ (see Eq.~(45)),
corresponding to solution (47), is equal to $\zeta \approx 0.00150$.
Note also that the position version of decomposition (45)
does not exist at $P=5$,
because then the number of unknowns is less than the number
of order equations, resulting in the absence of solutions.

As has been shown in the preceding subsection for the case of fourth-order
integration, algorithms with minimal numbers $n_{\rm f}$ of force
evaluation may not lead to optimal solutions. The reason is that slight
increasing $n_{\rm f}$ may significantly decrease the local errors and
thus overcompensate an increased computational efforts. So that increasing
$n_{\rm f}$ as well as $P$ on unity (note that $n_{\rm f}=P-1$)
and do not changing $n_{\rm g}$, i.e.
choosing $P=6$ with $n_{\rm f}=5$ and $n_{\rm g}=3$, it is possible to
derive from decomposition (13) up four (two velocity- and two position-like)
extended sixth-order schemes. They are
\begin{eqnarray}
{\rm CACABABACAC} \, , \ \ \ \ \ \ {\rm CABACACABAC} \, ,
\nonumber \\ [-7pt] \\ [-7pt]
{\rm ABACACACABA} \, , \ \ \ \ \ \ {\rm ACABACABACA} \, ,
\nonumber
\end{eqnarray}
where we have used an abbreviation that A and B denote exponential
operators ${\rm e}^{{\cal A} a_i \Delta t}$ and ${\rm e}^{{\cal B} b_i
\Delta t}$, respectively, whereas letter C corresponds to ${\rm
e}^{{\cal B} b_i \Delta t + {\cal C} c_i\Delta t^3}$. Each of these extended
schemes has itself correspondingly six, eight, four, and two sets of real
solutions for time coefficients. We have realized that the smallest values
of the norm $\zeta$ (see Eq.~(48)) of local errors ${\cal O}(\Delta t^7)$
within the sets are 0.0000264, 0.0000147, 0.000146, and
0.00000607, respectively. So that the last scheme should be considered as
the best. More explicit form for it is
\begin{eqnarray}
{\rm e}^{({\cal A}+{\cal B}) \Delta t +{\cal O}(\Delta t^7)} =
{\rm e}^{{\cal A} \rho \Delta t}
{\rm e}^{{\cal B} \vartheta \Delta t + \mu {\cal C} \Delta t^3}
{\rm e}^{{\cal A} \theta \Delta t}
{\rm e}^{{\cal B} \lambda \Delta t} &&
\nonumber \\
\times
{\rm e}^{{\cal A} \big(1 - 2 (\theta+\rho) \big) \frac{\Delta t}{2}}
{\rm e}^{{\cal B} \big(1 - 2 (\lambda+\vartheta) \big) \Delta t +
\chi {\cal C} \Delta t^3}
&&
\\
\times
{\rm e}^{{\cal A} \big(1 - 2 (\theta+\rho) \big) \frac{\Delta t}{2}}
{\rm e}^{{\cal B} \lambda \Delta t}
{\rm e}^{{\cal A} \theta \Delta t}
{\rm e}^{{\cal B} \vartheta \Delta t + \mu {\cal C} \Delta t^3}
{\rm e}^{{\cal A} \rho \Delta t} &&
\nonumber
\end{eqnarray}
with the optimal solution
\begin{eqnarray}
\rho      =  \ \ 0&.&1097059723948682{\rm E}\!+\!00
\nonumber \\
\theta    =  \ \ 0&.&4140632267310831{\rm E}\!+\!00
\nonumber \\
\vartheta =  \ \ 0&.&2693315848935301{\rm E}\!+\!00
\nonumber \\ [-7pt] \\ [-7pt]
\lambda  =   \ \ 0&.&1131980348651556{\rm E}\!+\!01
\nonumber \\
\chi    = \!    -0&.&1324638643416052{\rm E}\!-\!01
\nonumber \\
\mu   =      \ \ 0&.&8642161339706166{\rm E}\!-\!03
\nonumber
\end{eqnarray}
corresponding to $\zeta=0.00000607$. In such a way, the error function
has been reduced more than in 200 times with respect to scheme (47) for
which $\zeta \approx 0.00150$.

\subsubsection{Force-gradient algorithms of order eight}

In the case when $K=8$
we must satisfy up eighteen order conditions, namely,
$\nu=1$, $\sigma=1$, $\alpha=0$, $\beta=0$, $\gamma_{1-4}=0$, and
$\zeta_{1-10}=0$. Taking into account the symmetry of time coefficients
$a_p$, $b_p$, and $c_p$, this can be achieved at least at $P=12$, i.e.,
using $S=2P-1=23$ single exponential operators. For $P=12$ the velocity-
and position-like force-gradient decomposition (13) transforms into the
schemes
\begin{eqnarray}
{\rm CACACACACACACACACACACAC} && \ \ \ \ \ \
\\
{\rm and}
\ \ \ \ \ \ \ \ \ \ \ \ \ \ \ \ \ \ \ \ \ \
\ \ \ \ \ \ \ \ \ \ \ \ \ \ \ \ \ \ \ \ \ \
\ \ \ \ \ \ \ \ \ \ \ \,
&&
\nonumber \\
{\rm ACACACACACACACACACACACA} && \, ,
\end{eqnarray}
respectively. The number of unknowns for both the schemes are also eighteen
and we can try to solve the system of order conditions with respect to these
unknowns.

It is worth remarking such a system appears to be very cumbersome for schemes
under consideration. For instance, the resulting non-linear equations of this
system being written explicitly in Mathematica create a file of 0.5 Mb in
length! In view of this, our attempts to solve the equations symbolically
have not meet with much success. We mention that all the results presented
above for algorithms of orders 2, 4, and 6 have been solved analytically
or in quadratures. Saying in quadratures we mean that the problem was
reduced to finding real zeros for a one-dimensional polynomial of a given
order. So that we could identify exactly the number of solutions and their
locations. Here the situation is somewhat different because we must solve
the system using purely numerical approaches, such as the Newton method.
As a result, one cannot guarantee that we will found all possible solutions.
However, solving the system on a computer during significantly long time,
one can stay with a great probability that we have found almost all physically
interesting solutions and chosen among them nearly optimal sets.

The numerical calculations has been performed in Fortran using the
well-recognized Newton solver with numerical determination of partial
derivatives. The values for non-linear functions (that constitute the
system
of equations) were obtained using recursive relations (20)--(27), (A1),
and (A2), but not explicit expressions for them to save the processor
time and increase the precision of the computations. The initial guess
for solutions were generated at random within the interval $[-2.5, 2.5]$
in each the eighteenth directions. If Newton's iterations become to
diverge at a particular guess or during the calculations, a next random
point was involved to repeat the process. In such a way, after several
days of continuous attacking the systems of equations on an Origin 3800
workstation, we found two and five solutions for schemes (52) and (53),
respectively. The optimal among them are following
\small
\begin{eqnarray}
a_1 = \ \ 0&&
\nonumber \\ [-2pt]
b_1 = b_{12} = \ \ 0&.&1839699354244402{\rm E}\!+\!00
\nonumber \\ [-2pt]
c_1 = c_{12} = \ \ 0&&
\nonumber \\ [-2pt]
a_2 = a_{12} = \ \ 0&.&6922517172738832{\rm E}\!+\!00
\nonumber \\ [-2pt]
b_2 = b_{11} = \ \ 0&.&7084389757230299{\rm E}\!+\!00
\nonumber \\ [-2pt]
c_2 = c_{11} = \ \ 0&.&3976209968238716{\rm E}\!-\!01
\nonumber \\ [-2pt]
a_3 = a_{11} =  -0&.&3183450347119991{\rm E}\!+\!00
\nonumber \\ [-2pt]
b_3 = b_{10} = \ \ 0&.&1981440445033534{\rm E}\!+\!00
\nonumber \\ [-2pt]
c_3 = c_{10} = \ \ 0&.&2245403440322733{\rm E}\!-\!01
\nonumber \\ [-2pt]
a_4 = a_{10} = \ \ 0&.&6766724088765565{\rm E}\!+\!00
\nonumber \\ [-2pt]
b_4 = b_9  =  -0&.&6409380745116974{\rm E}\!-\!01
\nonumber \\ [-2pt]
c_4 = c_9  = \ \ 0&.&9405266232181224{\rm E}\!-\!03
\nonumber \\ [-2pt]
a_5 = a_9  =  -0&.&7207972470858706{\rm E}\!+\!00
\nonumber \\ [-2pt]
b_5 = b_8  =  -0&.&6887429532761409{\rm E}\!+\!00
\nonumber \\ [-2pt]
c_5 = c_8  =  -0&.&7336500519635302{\rm E}\!-\!01
\nonumber \\ [-2pt]
a_6 = a_8  = \ \ 0&.&3580316862350045{\rm E}\!+\!00
\nonumber \\ [-2pt]
b_6 = b_7  = \ \ 0&.&1622838050764871{\rm E}\!+\!00
\nonumber \\ [-2pt]
c_6 = c_7  = \ \ 0&.&2225664796363730{\rm E}\!-\!01
\nonumber \\ [-2pt]
a_7 =         -0&.&3756270611751488{\rm E}\!+\!00
\nonumber
\end{eqnarray}
\normalsize
for velocity-like integration (52), and
\small
\begin{eqnarray}
      b_{12} = \ \ 0 &&
\nonumber \\ [-2pt]
      c_{12} = \ \ 0 &&
\nonumber \\ [-2pt]
a_1 = a_{12} =   \ \ 0&.&41009674738801111928784693005080{\rm E}\!+\!00
\nonumber \\ [-2pt]
b_1 = b_{11} =   \ \ 0&.&48249309817414952912695842664785{\rm E}\!-\!02
\nonumber \\ [-2pt]
c_1 = c_{11} =   \ \ 0&.&14743936907797528364717244760736{\rm E}\!-\!03
\nonumber \\ [-2pt]
a_2 = a_{11} =  -0&.&34123345756052780489101697378499{\rm E}\!+\!00
\nonumber \\ [-2pt]
b_2 = b_{10} =   \ \ 0&.&17492394861090375603419001374207{\rm E}\!+\!00
\nonumber \\ [-2pt]
c_2 = c_{10} =   \ \ 0&.&23288450531932545357194967600155{\rm E}\!-\!03
\nonumber \\ [-2pt]
a_3 = a_{10} =   \ \ 0&.&25644714021068150492361761631743{\rm E}\!+\!00
\nonumber \\ [-2pt]
b_3 = b_9  =   \ \ 0&.&29304366370957066164364546204288{\rm E}\!+\!00
\nonumber \\ [-2pt]
c_3 = c_9  =   \ \ 0&.&61648659635535962497705619884752{\rm E}\!-\!02
\nonumber \\ [-2pt]
a_4 = a_9  =   \ \ 0&.&27765273975812438394100476242641{\rm E}\!+\!00
\nonumber \\ [-2pt]
b_4 = b_8  =   \ \ 0&.&47448940168459770284238136482511{\rm E}\!-\!01
\nonumber \\ [-2pt]
c_4 = c_8  =  -0&.&12307516860831240716732016960034{\rm E}\!-\!01
\nonumber \\ [-2pt]
a_5 = a_8  =  -0&.&56926266869753773902939657321159{\rm E}\!+\!00
\nonumber \\ [-2pt]
b_5 = b_7  =  -0&.&15299863411743974499219652320477{\rm E}\!-\!02
\nonumber \\ [-2pt]
c_5 = c_7  =  -0&.&73296648559126385387017161643798{\rm E}\!-\!04
\nonumber \\ [-2pt]
a_6 = a_7  =   \ \ 0&.&46629949890124853576794423820194{\rm E}\!+\!00
\nonumber \\ [-2pt]
b_6        =  -0&.&37422994259002571606842462603791{\rm E}\!-\!01
\nonumber \\ [-2pt]
c_6        =   \ \ 0&.&15295860994523744731993293847001{\rm E}\!-\!01
\nonumber
\end{eqnarray}
\normalsize
for its position-like counterpart (53). The number of force evaluations
per times step for schemes (52) and (53) is $n_{\rm f}=P-1=11$, whereas
the number of force-gradient recalculations consists $n_{\rm g}=10$
(since $c_1=0$ and thus the two boundary letters C in formula (52) should
be actually replaced by B) and $n_{\rm g}=11$, respectively.

In view of a complicated structure of the ninth-order truncation
uncertainties ${\cal O}(\Delta t^9)$, the optimal solutions just
presented have been chosen in somewhat other way than above, namely, by
providing a minimum for the function $\delta=\max_{p=1}^{P}(|a_p|,|b_p|)$.
This simplified criterion was used, in particular, by Kahan and Li
\cite{Lidis,KahanLi}, when optimizing usual force algorithms. As a
result, we have obtained $\delta_{\rm min} \equiv |a_5| = |a_9|
\approx 0.721$ for scheme (52) and $\delta_{\rm min} \equiv |a_5| =
|a_8| \approx 0.569$ for scheme (53). Since $\delta_{\rm min}$ is smaller
in the last case, the position-like integration should be considered as more
preferable. Its time coefficients have been presented even with thirty
second significant digits to be used in applications for very
accurate integration. In order to ensure that all the digits shown are
correct in both the cases, we have carried out a few additional Newton's
iterations in Maple with up 200 digits during the internal computations,
and taking as initial guesses the solutions already obtained in Fortran.

The position-like decomposition (53) has also another advantage over
the velocity version (52) in that the all the intermediate ($q \le P$)
states in velocity and position space stay during the integration within
a given interval $[0,\Delta t]$, i.e., $0 \le \sum_{p=1}^q a_p \le 1$ and
$0 \le \sum_{p=1}^q b_p \le 1$. This property may be important when
solving ordinary differential equations (for specific physical systems
or in pure mathematics applications) with singularities beyond the
interval of the integration (note that, in particular, any system of
differential equations of the form ${\rm d}^2 {\bf x}/{\rm d} t^2 =
{\bf f}({\bf x})$ is reduced to the equations of motion under
consideration in this paper).
Note also that in order to construct eighth-order schemes within usual
decomposition approach (6) (i.e., without involving any force-gradients),
it could be necessary to apply up $2 \cdot 18-1=35$ (instead of 23) single
exponential propagators. Such schemes has never been derived by
decomposition (6) because of the serious technical difficulties. They can
be explicitly introduced only by compositions of lower-order integrators
(see the next section). Instead, using generalized scheme (13) has allowed
us to derive eighth-order algorithms by direct decompositions for the first
time (the force-gradient algorithms presented in subsection II.B.{\em
5} for order six are completely new as well).

All the decomposition algorithms obtained by us in subsections II.B.{\em 3},
{\em 4}, {\em 5}, and {\em 6} are collected below in Table~1. Here, the
designations Err3, Err5, and Err7 relate to the norms $\sqrt{\alpha^2+
\beta^2}$, $\gamma$, and $\zeta$ of correspondingly third-, fifth-, and
seventh-order truncation errors (see Eqs.~(6), (15)--(18), (34), and (48)),
whereas $n_{\rm f}$ and $n_{\rm f}$ denote the numbers of force and
force-gradient evaluations per time step. The optimal algorithms for
orders 2, 4, 6, and 8 are labeled by G2, C$'$, G6, and G8, respectively.
Among other schemes presented for each given order, such algorithms reduce
the truncation uncertainties to a minimum. Taking into account that this
reduction is achieved at the same or nearly the same computational efforts,
the optimal algorithms should be considered as the best not only with
respect to their precision but in view of the overall efficiency as well
(see also comments on this in section III).

\end{multicols}

\begin{table}
\caption{The basic decomposition force-gradient algorithms}

\vspace{6pt}

\begin{tabular}{lccccccccl}
\ \ Algorithm & \ \ Order \ \ & \ $n_{\rm f}$ \
& \ $n_{\rm g}$ \ & \ \ \ \ Err3 \ \ \ \ & \ \ \ \ Err5 \ \ \ \
& \ \ \ \ Err7 \ \ \ \ & Equations & Remarks & $\!\!\!\!\!\!$ Label \\
\hline
\hline \\ [-8pt]
\ \ \ \ CAC           & 2 &  1    &  1    & $8.33 \!\cdot\! 10^{-2}$ & $1.34 \!\cdot\! 10^{-2}$ &  $2.24 \!\cdot\! 10^{-3}$ &    (28)       &     New                 &          G2$'$    \\
\ \ \ \ ACA           & 2 &  1    &  1    & $4.17 \!\cdot\! 10^{-2}$ & $6.48 \!\cdot\! 10^{-3}$ &  $7.25 \!\cdot\! 10^{-4}$ &    (29)       & \ \ New$^{+}$           &          G2       \\
\hline \\ [-8pt]
\ \ \ \ BACAB         & 4 &  2    &  1    &            0             & $7.13 \!\cdot\! 10^{-4}$ &  $6.30 \!\cdot\! 10^{-5}$ &  (30,32)      &     Refs.~\cite{Suzuki,Chin} \,    & \        A        \\
\ \ \ \ CABAC         & 4 &  2    &  1    &            0             & $3.34 \!\cdot\! 10^{-3}$ &  $2.72 \!\cdot\! 10^{-4}$ &  (30,33)      &     New                 & \        A$'$     \\
\ \ \ \ CACAC         & 4 &  2    &  2    &            0             & $5.95 \!\cdot\! 10^{-4}$ &  $4.83 \!\cdot\! 10^{-5}$ &  (30,35)      &     New                 & \        A$''$    \\
\ \ \ \ ACACA         & 4 &  2    &  2    &            0             & $7.15 \!\cdot\! 10^{-4}$ &  $5.59 \!\cdot\! 10^{-5}$ &  (36,37)      &     Refs.~\cite{Suzuki,Chin} \,    & \        B        \\
\ \ \ \ ABACABA       & 4 &  3    &  1    &            0             & \ \ \ ${1.41 \!\cdot\! 10^{-4}}^{\rm \! (a)}$ & \ \ \ ${1.04 \!\cdot\! 10^{-5}}^{\rm \! (a)}$ &  (38,41/42)   & \ \ Ref.~\cite{Chin}/New$^{+}$ & $\!\!\!$ C/C$'$ \\
\ \ \ \ CABABAC       & 4 &  3    &  1    &            0             & \ \ \ ${8.55 \!\cdot\! 10^{-4}}^{\rm \! (b)}$ & \ \ \ ${2.24 \!\cdot\! 10^{-5}}^{\rm \! (b)}$ &  (39,43/44)   &     Ref.~\cite{ChinChen}/New       & $\!\!\!$ D/D$'$ \\
\hline \\ [-8pt]
\ \ \ \ BACACACAB     & 6 &  4    &  3    &            0             &            0             &  $1.50 \!\cdot\! 10^{-3}$ &  (45,47)      &     New                 & \        G6$'$    \\
\ \ \ \ CACABABACAC   & 6 &  5    &  3    &            0             &            0             &  $2.64 \!\cdot\! 10^{-5}$ &    (49)       &     New                 & \        G6$''$   \\
\ \ \ \ CABACACABAC   & 6 &  5    &  3    &            0             &            0             &  $1.47 \!\cdot\! 10^{-5}$ &    (49)       &     New                 & \        G6$'''$  \\
\ \ \ \ ABACACACABA   & 6 &  5    &  3    &            0             &            0             &  $1.46 \!\cdot\! 10^{-4}$ &    (49)       &     New                 & \        G6$''''$ \\
\ \ \ \ ACABACABACA   & 6 &  5    &  3    &            0             &            0             &  $6.07 \!\cdot\! 10^{-6}$ &  (50,51)      & \ \ New$^{+}$           & \        G6       \\
\hline \\ [-8pt]
\ \ \ \ {\footnotesize
BACACACACACA} \\ [-6.5pt]
                      & 8 & 11 \, & 10 \, &            0             &            0             &            0          &        (52)       &     New                 & \        G8$'$    \\ [-6.5pt]
\ \ \ \ $\times${\footnotesize
CACACACACAB} \\ [2.5pt]
\ \ \ \ {\footnotesize
ACACACACACAC} \\ [-6.5pt]
                      & 8 & 11 \, & 11 \, &            0             &            0             &            0          &        (53)       & \ \ New$^{+}$           & \        G8       \\ [-6.5pt]
\ \ \ \ $\times${\footnotesize
ACACACACACA}
\end{tabular}

\vspace{6pt}

{\small
$^+$ The best algorithm within a given order}

{\small
$^{\rm (a)}$The value corresponding to scheme C$'$}

{\small
$^{\rm (b)}$The value corresponding to scheme D$'$}
\end{table}

\begin{multicols}{2}

Finally, it is worth remarking that the problem of constructing algorithms
with only positive coefficients $a_p$ and $b_p$ for orders six and higher
still remains. We mention that for order four, this problem has been
resolved (see subsection II.B.{\em 4}) by transferring the force-gradient
component of truncation uncertainties into the exponential propagators.
For orders $K \ge 6$, additional higher-order gradients should appear under
these exponentials to provide the required positiveness. Our analysis has
shown, however, that such high-order exponentials (besides their very
cumbersome forms) cannot be evaluated in quadratures and need in performing
implicit calculations by iteration. In view of this we can come to a
conclusion that beyond fourth order,
analytically integrable decomposition algorithms with purely positive
coefficients do not exist. Mathematically rigorous prove of this
statement will be considered in our further investigation and presented
elsewhere.

\subsection{Integration by advanced compositions}

With increasing the order of integration to ten and higher, the construction
of algorithms by direct decompositions (13) becomes to be inefficient
because of a large number of the order conditions and time coefficients.
However, having the already derived force-gradient integrators of lower
orders $K$, we can try to compose them as
\begin{equation}
S_Q(\Delta t) = S_K(d_1 \Delta t) \ldots S_K(d_P \Delta t) \ldots
                S_K(d_1 \Delta t)
\end{equation}
for obtaining an algorithm of order $Q > K$. Then the composition constants
$d_p$, where $p=1,2,\ldots,P$, should be chosen in such a way to provide
the maximal possible value of $Q$ at a given number $P \ge 2$. Note that
lower-order propagations $S_K(d_p \Delta t)$ enter symmetrically in
composition (54) and their total number $2P-1$ accepts odd values. So
that if a basic integrator $S_K$ is self-adjoint, the resulting
algorithm $S_Q$ will be self-adjoint as well. The
idea of using formula like (54) is not new and has been applied
by different authors in previous investigations \cite{Yoshida,Suzukip,%
Suzukium,Lidis,Qin,McLachlan,KahanLi,Murua}. But these investigations
were focused, in fact, on the compositions of usual second-order ($K=2$)
schemes (to our knowledge, no actual calculations of composition
constants for fourth- and higher-order-based integrators
have been reported). Although using the second-order-based approach
allowed to introduce algorithms to the tenth
order \cite{Suzukium,Lidis,KahanLi}, further increasing $Q$ has led to
unresolved numerical difficulties when finding the coefficients of the
compositions.

Usually, these difficulties are obviated with the help of Creutz's and
Gocksch's method \cite{Creutz}. We mention that
according to this method, an algorithm
of order
$K+2$ can be derived by the triplet concatenation
\begin{equation}
S_{K+2}(\Delta t) = S_K(D_K \Delta t) S_K \big( (1-2 D_K) \Delta t \big)
                    S_K(D_K \Delta t)
\end{equation}
of a self-adjoint integrator of order $K$, where $D_K=1/(2-2^{1/(K+1)})$.
In particular, Chin and Kidwell \cite{Chins} starting from force-gradient
algorithm (41) of order four and repeating procedure (55) up to order 12,
have indicated a visible increasing the efficiency of the computation
with respect to second-order-based schemes. In this approach, however,
the number of force and force-gradient evaluations (the most time-consuming
part of the calculations) increases too rapidly with increasing $K$,
namely as $3^{(K-4)/2}$ relatively to the fourth-order integrator.

The present study is aimed to overcome the above problems by an explicit
consideration of four-, sixth-, and eighth-order-based (force-gradient)
algorithms within general composition approach (54). This results in
reducing the total number of basic propagations
to a minimum and providing significant speeding
up the integration. The composition algorithms are derived up to the
sixteenth order inclusively.

\subsubsection{Fourth-order based algorithms}

In the case when $K=4$, the basic self-adjoint propagation is
\begin{equation}
S_4(\tau) = {\rm e}^{{\cal X}_1 \tau + {\cal X}_5 \tau^5 +
            {\cal X}_7 \tau^7 + {\cal X}_9 \tau^9 +
            {\cal X}_{11} \tau^{11} + \ldots} \, ,
\end{equation}
where ${\cal X}_1 \equiv {\cal A} + {\cal B}$. Explicit form of higher-order
truncation operators ${\cal X}_5$, ${\cal X}_7$, ${\cal X}_9$, ${\cal X}_{11}$,
and so on (which was previously found for ${\cal X}_5$ and ${\cal X}_7$,
see Eqs.~(17) and (18)) are not important within the composition approach.
Then formula (54) reduces to series ($n=0,1,\ldots,P-2$) of the transformation
\begin{equation}
S_Q^{(n+1)}(\Delta t) = S_4(d^{(n)} \Delta t) S_Q^{(n)}(\Delta t)
                        S_4(d^{(n)} \Delta t)
\end{equation}
with $S_Q^{(0)}=S_4(d_P \Delta t)$ and $d^{(n)}=d_{P-n-1}$. In view of
Eqs.~(56) and (57), the structure of resulting propagation can be cast
at each $n$ as
\begin{equation}
S_Q(\Delta t) \!=\! {\rm e}^{{\cal Y}_1 \Delta t + {\cal Y}_5 \Delta t^5 \!+\!
                {\cal Y}_7 \Delta t^7 \!+\! {\cal Y}_9 \Delta t^9 \!+\!
                {\cal Y}_{11} \Delta t^{11} + {\cal O}(\Delta t^{13})} ,
\end{equation}
with
$$
{\cal Y}_1 = q_1 {\cal X}_1 , \ \
{\cal Y}_5 = q_2 {\cal X}_5 , \ \
{\cal Y}_7 = q_3 {\cal X}_7 + q_4 [{\cal X}_1,{\cal X}_1,{\cal X}_5] ,
$$
\begin{equation}
{\cal Y}_9 \!=\! q_5 {\cal X}_9 \!+\!
q_6 [{\cal X}_1,{\cal X}_1,{\cal X}_7] \!+\!
q_7 [{\cal X}_1,{\cal X}_1,{\cal X}_1,{\cal X}_1,{\cal X}_5] ,
\end{equation}
\begin{eqnarray*}
\\ [-26pt]
{\cal Y}_{11} =
&&
q_8 {\cal X}_{11} + q_9 [{\cal X}_1,{\cal X}_1,{\cal X}_9] +
q_{10} [{\cal X}_1,{\cal X}_1,{\cal X}_1,{\cal X}_1,{\cal X}_7] +
\\ &&
q_{11} [{\cal X}_1,{\cal X}_1,{\cal X}_1,{\cal X}_1,
{\cal X}_1,{\cal X}_1,{\cal X}_5] +
q_{12} [{\cal X}_5,{\cal X}_1,{\cal X}_5] .
\end{eqnarray*}
Comparing (56) and (58) yields values of $q$-multipliers at $n=0$, namely,
$q_1^{(0)}=d_P$, $q_2^{(0)}=d_P^{K+1}$, $q_3^{(0)}=d_P^{K+3}$, $q_5^{(0)}=
d_P^{K+5}$, and $q_8^{(0)}=d_P^{K+7}$, whereas, $q_4^{(0)}=q_6^{(0)}=
q_7^{(0)}=q_9^{(0)}=q_{10}^{(0)}=q_{11}^{(0)}=q_{12}^{(0)}=0$. Expanding
both the sides of Eq.~(57) into Taylor's series with respect to $\Delta t$,
one finds that values for these multipliers at $n>0$ can be obtained using
the following recursive relations
\end{multicols}
$$
q_1^{(n+1)} = q_1^{(n)} + 2 d^{(n)} \, , \ \ \ \ \
q_2^{(n+1)} = q_2^{(n)} + 2 {d^{(n)}}^{K+1} \, ,
$$

\vspace{0pt}

$$
q_3^{(n+1)} = q_3^{(n)} + 2 {d^{(n)}}^{K+3} \, , \ \ \ \ \ \ \
q_4^{(n+1)} = q_4^{(n)} + d^{(n)} \big(q_1^{(n)} +
              d^{(n)}\big) \big(q_1^{(n)} {d^{(n)}}^K - q_2^{(n)}\big)/6
\, ,
$$

\vspace{0pt}

$$
q_5^{(n+1)} = q_5^{(n)} + 2 {d^{(n)}}^{K+5} \, , \ \ \ \ \ \ \
q_6^{(n+1)} = q_6^{(n)} + d^{(n)} \big(q_1^{(n)} +
              d^{(n)}\big) \big(q_1^{(n)} {d^{(n)}}^{K+2} - q_3^{(n)}\big)/6
\, ,
$$

\vspace{4pt}

\begin{eqnarray*}
q_7^{(n+1)}=&&q_7^{(n)} + d^{(n)} \big(q_1^{(n)}+
d^{(n)}\big) \big( {q_1^{(n)}}^2 q_2^{(n)}-
60 q_4^{(n)}+7 q_1^{(n)} q_2^{(n)} d^{(n)}+
\\ [6pt] &&
7 q_2^{(n)} {d^{(n)}}^2-
{q_1^{(n)}}^3 {d^{(n)}}^K-
7 {q_1^{(n)}}^2 {d^{(n)}}^{K+1}-7 q_1^{(n)} {d^{(n)}}^{K+2}\big)/360
\, ,
\end{eqnarray*}

\vspace{0pt}

$$
q_8^{(n+1)} = q_8^{(n)} + 2 {d^{(n)}}^{K+7} \, , \ \ \ \ \ \ \
q_9^{(n+1)} = q_9^{(n)} + d^{(n)} \big(q_1^{(n)} +
              d^{(n)}\big) \big(q_1^{(n)} {d^{(n)}}^{K+4} - q_5^{(n)}\big)/6
\, ,
$$

\vspace{-10pt}

\begin{equation}
\end{equation}

\vspace{-10pt}

\begin{eqnarray*}
q_{10}^{(n+1)}=&&q_{10}^{(n)}+
d^{(n)} \big(q_1^{(n)}+d^{(n)}\big) \big({q_1^{(n)}}^2 q_3^{(n)}-60 q_6^{(n)}+
7 q_1^{(n)} q_3^{(n)} d^{(n)}+
\\ [6pt] &&
7 q_3^{(n)} {d^{(n)}}^2-{q_1^{(n)}}^3 {d^{(n)}}^{K+2}-
7 {q_1^{(n)}}^2 {d^{(n)}}^{K+3} - 7 q_1^{(n)} {d^{(n)}}^{K+4}\big)/360
\, ,
\end{eqnarray*}

\vspace{4pt}

\begin{eqnarray*}
q_{11}^{(n+1)}=&&q_{11}^{(n)} +
d^{(n)} \big(q_1^{(n)} + d^{(n)}\big) \big(
42 {q_1^{(n)}}^2 q_4^{(n)} - {q_1^{(n)}}^4 q_2^{(n)}
- 2520 q_7^{(n)} - 11 {q_1^{(n)}}^3 q_2^{(n)} d^{(n)} +
294 q_1^{(n)} q_4^{(n)} d^{(n)} -
\\ [6pt] &&
42 {q_1^{(n)}}^2 q_2^{(n)} {d^{(n)}}^2 +
294 q_4^{(n)} {d^{(n)}}^2 -
62 q_1^{(n)} q_2^{(n)} {d^{(n)}}^3 +
{q_1^{(n)}}^5 {d^{(n)}}^K - 31 q_2^{(n)} {d^{(n)}}^4 +
\\ [6pt] &&
11 {q_1^{(n)}}^4 {d^{(n)}}^{K+1} +
42 {q_1^{(n)}}^3 {d^{(n)}}^{K+2} +
62 {q_1^{(n)}}^2 {d^{(n)}}^{K+3} + 31 q_1^{(n)} {d^{(n)}}^{K+4}\big)/15120
\, ,
\end{eqnarray*}

\vspace{-2pt}

$$
q_{12}^{(n+1)}=q_{12}^{(n)}+d^{(n)} \big(q_1^{(n)} {d^{(n)}}^4-
q_2^{(n)}\big) \big(q_2^{(n)}+{d^{(n)}}^5\big)/6
\, .
$$

\vspace{4.5pt}

\begin{multicols}{2}

Applying the above relations $P-1$ times will give the final values of
$q$-multipliers and thus lead to the desired order conditions. For
instance, the first condition is very simple and reads: $q_1 = d_P +
2 \sum_{p=1}^{P-1} d_p = 1$. This provides ${\cal Y}_1={\cal X}_1$ and
guarantees (see Eqs.~(56), (58) and (59)) that the order of the composition
scheme will be at least not lower than that of the basic scheme, i.e. $Q
\ge 4$ in our case. All other multipliers $q_2$, $q_3$, $q_4$, \ldots
$q_{\cal N}$
should be consecutively set to zero, forming higher-order conditions. The
total number ${\cal N}$ of the conditions depends on a required order $Q>4$
of the composition scheme. In particular, at $Q=6$ we must kill the term
${\cal Y}_5$ at fifth-order truncation uncertainties (see Eq.~(58)). Taking
into account Eq.~(59), this results in two order conditions, namely, $q_1=1$
and $q_2=0$ which can be satisfied at $P=2$. Then one obtains a system of
equations, $q_1=2d_1+d_2=0$, and $q_1=2d_1^5+d_2^5=0$, with respect
to two unknowns, $d_1$ and $d_2$. The system can be solved analytically,
and the solution is $d_1=1/(2-2^{1/5}) \equiv D_4$ with $d_2=1-2d_1$ that
coincides (at $K=4$) with the result of triplet construction (55).
This coinciding is not surprising because, as can be seen easily, both
approaches (54) and (55) are identical in a partial case when $P=2$
and $Q-K=2$.

With further increasing $Q$, composition approach (54) will lead to a
more efficient integration. Indeed, choosing $Q=8$ requires the term
${\cal Y}_7$ in Eq.~(58) should be killed additionally. This is achieved
by putting $q_3=q_4=0$ in Eq.~(59), and, therefore, by solving at $P=4$
a system of four non-linear equations, $q_1=1$, $q_2=0$, $q_3=0$, and
$q_4=0$ with respect to the same number of unknowns $d_1$, $d_2$, $d_3$,
and $d_4$. So that the minimal number of fourth-order integrators in the
eight-order composition should be $2P-1=7$, whereas this number is equal
to $3^{(Q-K)/2}=9$ when triplet concatenation (55) is used. Expressions
for the non-linear equations can readily be reproduced by applying
the corresponding set of recursive relations (60). We will not present
such expressions explicitly, because as has been realized, the order
equations do not allow to be solved analytically at $Q-K \ge 4$ for
any $K \ge 4$.
But, these equations can be solved
in a quite efficient way numerically using the Newton's method. Details of
the numerical calculations are similar to those described in subsection
II.B.{\em 6}. Here (when $P=4$, $K=4$, and $Q=8$) we have found five
solutions, and it seems no other real solutions exist. The optimal set
is
\begin{eqnarray}
    d_1 =  \ \ 0&.&8461211474696757{\rm E}\!+\!00
\nonumber \\
    d_2 =  \ \ 0&.&1580128458008567{\rm E}\!+\!00
\nonumber \\ [-7pt] \\ [-7pt]
    d_3 =     -0&.&1090206660543938{\rm E}\!+\!01
\nonumber \\
    d_4 =  \ \ 0&.&1172145334546811{\rm E}\!+\!01 \, .
\nonumber
\end{eqnarray}
Solution (61) simultaneously leads the smallest values for the maximal
composition coefficient $\max_{p=1}^4{|d_p|} \approx 1.172$ and the norm
$(q_5^2+q_6^2+q_7^2)^{1/2} \approx 0.270$ of the main ninth-order term
${\cal Y}(\Delta t^9)$ of truncations uncertainties.

When deriving tenth-order composition algorithms (at $K=4$), i.e. when
$Q=10$,
three additional order conditions arise, $q_5=0$, $q_6=0$, and $q_7=0$,
needed
to eliminate the term ${\cal Y}(\Delta t^9)$
(see Eqs.~(58) and (59)).
Then we come in
overall to 7 non-linear equations which can be satisfied by appropriate
choosing composition constants $d_p$ ($p=1,2,\ldots,P$) at $P=7$. In
this case, we have identified more than 150 real solutions and probably
there are somebody others (we stopped the searching after several days
of the computations). Among the solutions found the optimal set looks

\small
\begin{eqnarray}
    d_1 = \ \ 0&.&80523995769578082326628169802782
\nonumber \\ [-2pt]
    d_2 =    -0&.&49193105914623101022388138864143
\nonumber \\ [-2pt]
    d_3 = \ \ 0&.&35449258654398460535529269988483
\nonumber \\ [-8pt] \\ [-8pt]
    d_4 =    -0&.&69573922271140223803036463461997
\nonumber \\ [-2pt]
    d_5 = \ \ 0&.&39959538030329256359349977087819
\nonumber \\ [-2pt]
    d_6 = \ \ 0&.&54979568601438452794128031563760
\nonumber
\end{eqnarray}
\normalsize
and $d_7=1-2 (d_1+d_2+d_3+d_4+d_5+d_6)$. This set minimizes at once
the norm $(q_8^2+q_9^2+q_{10}^2+q_{11}^2+q_{12}^2)^{1/2}$ of the main
eleventh-order term ${\cal Y}(\Delta t^{11})$ of truncation errors
and the quantity $\max_{p=1}^7{|d_p|}$ to the values 0.00412 and
0.843 ($\equiv |d_7|$), respectively. Here, the number of basic
propagations (stages)
is $2P-1=13$, i.e. more than in two times smaller than this
number $3^{(Q-K)/2}=27$ within triplet concatenation (55).

In order to introduce twelfth-order algorithms, $Q=12$, on the basis
of fourth-order compositions it is necessary to deal with $P=12$
unknowns $d_p$ to fulfill the same number of the order conditions
$q_1=1$, and $q_{2-12}$=0. Here we have found more than 200 real
solutions and perhaps there are somebody else. The best among
them, which minimizes $\max_{p=1}^{12}{|d_p|}$ to the value 0.611
($\equiv |d_{12}|$), is
\small
\begin{eqnarray}
    d_1    = \ \ 0&.&17385016093097855436061712858303
\nonumber \\ [-2pt]
    d_2    = \ \ 0&.&53377479890712207949282653990842
\nonumber \\ [-2pt]
    d_3    = \ \ 0&.&12130138614668307673802291966495
\nonumber \\ [-2pt]
    d_4    = \ \ 0&.&29650747033807195273440032505629
\nonumber \\ [-2pt]
    d_5    =    -0&.&59965999857335454018482312008233
\nonumber \\ [-8pt] \\ [-8pt]
    d_6    = \ \ 0&.&09043581286204437145871130429094 \
\nonumber \\ [-2pt]
    d_7    =    -0&.&43979146257635806886778748138962
\nonumber \\ [-2pt]
    d_8    =    -0&.&30251552922346495057010240779104
\nonumber \\ [-2pt]
    d_9    = \ \ 0&.&59895872989247982114545906953712
\nonumber \\ [-2pt]
    d_{10} = \ \ 0&.&31236416538275576151816280776696
\nonumber \\ [-2pt]
    d_{11} =    -0&.&59081230769647833184090443445303
\nonumber
\end{eqnarray}
\normalsize
with $d_{12}=1-2 \sum_{p=1}^{11} d_p$. Thus,
the minimal
number of fourth-order stages needed to compose the twelfth-order algorithm
is $2P-1=23$, instead of up $3^{(Q-K)/2}=81$ as in the case of usual triplet
construction (55).

\subsubsection{Sixth- and eighth-order based algorithms}

When $K=6$ or 8, the basic propagation reads
\begin{equation}
S_6(\tau) = {\rm e}^{{\cal X}_1 \tau + {\cal X}_7 \tau^7 +
            {\cal X}_9 \tau^9 + {\cal X}_{11} \tau^{11} +
            {\cal X}_{13} \tau^{13} + \ldots} \, ,
\end{equation}
or
\begin{equation}
S_8(\tau) = {\rm e}^{{\cal X}_1 \tau + {\cal X}_9 \tau^9 +
            {\cal X}_{11} \tau^{11} + {\cal X}_{13} \tau^{13} +
            {\cal X}_{15} \tau^{15} + \ldots} \, ,
\end{equation}
respectively. Here, the compositions reduce to the recursive
transformation
\begin{equation}
S_Q^{(n+1)}(\Delta t) = S_{6,8}(d^{(n)} \Delta t) S_Q^{(n)}(\Delta t)
                        S_{6,8}(d^{(n)} \Delta t)
\end{equation}
with $S_Q^{(0)}$ being equal to $S_6(d_P \Delta t)$ or $S_8(d_P \Delta t)$
and $n=0,1,\ldots,P-2$. The left-hand-side of expression (66) can again be
presented at each $n$ as a single exponential,
$$
S_Q(\Delta t) = {\rm e}^{{\cal Y}_1 \tau + {\cal Y}_{K\!+\!1} \tau^{K\!+\!1} +
                {\cal Y}_{K\!+\!3} \tau^{K\!+\!3} +
                {\cal Y}_{K\!+\!5} \tau^{K\!+\!5} +
                {\cal Y}_{K\!+\!7} \tau^{K\!+\!7} + \ldots} ,
$$
where now
\begin{eqnarray}
{\cal Y}_1 &=& q_1 {\cal X}_1 , \ \ \
{\cal Y}_{K+1} = q_2 {\cal X}_{K+1} ,
\nonumber \\ [6pt]
{\cal Y}_{K+3} &=& q_3 {\cal X}_{K+3} +
q_4 [{\cal X}_1,{\cal X}_1,{\cal X}_{K+1}] \, ,
\nonumber \\ [6pt]
{\cal Y}_{K+5} &=& q_5 {\cal X}_{K+5} +
q_6 [{\cal X}_1,{\cal X}_1,{\cal X}_{K+3}] +
\nonumber \\ [-6pt] \\ [-6pt]
&&
q_7 [{\cal X}_1,{\cal X}_1,{\cal X}_1,{\cal X}_1,{\cal X}_{K+1}] ,
\nonumber \\ [6pt]
{\cal Y}_{K+7} &=&
q_8 {\cal X}_{K+7} + q_9 [{\cal X}_1,{\cal X}_1,{\cal X}_{K+5}] +
\nonumber \\ [1pt] &&
q_{10} [{\cal X}_1,{\cal X}_1,{\cal X}_1,{\cal X}_1,{\cal X}_{K+3}] +
\nonumber \\ [1pt] &&
q_{11} [{\cal X}_1,{\cal X}_1,{\cal X}_1,{\cal X}_1,
{\cal X}_1,{\cal X}_1,{\cal X}_{K+1}] \, .
\nonumber
\end{eqnarray}
Recursive relations for multipliers $q_{1-11}$, corresponding to
transformation (66), remain the same in form as in the case $K=2$.
So that we should merely to put either $K=6$ or $K=8$ in Eq.~(60)
to obtain the required set of order conditions.

In view of the equivalence of Eqs.~(54) and (55) at $Q=K+2$, the first
step on increasing the order of composition scheme to $Q=8$ when $K=6$
or $Q=10$ when $K=8$ is trivial and yields $P=2$, $d_1=1/(2-2^{1/(K+1)})
\equiv D_K$, and $d_2=1-2d_1$. The next steps on increasing $Q$ to the
higher values
 $K+4$, $K+6$, and $K+8$ at $K=6$ or 8 are similar
to the steps described above for $K=2$. Namely, they lead to the necessity
of solving numerically the system of $P$ non-linear equations, $q_1=1$,
$q_2=0$, \ldots, $q_P=0$, with
$P=4$, 7, and 11, respectively. The only difference from the case $K=2$ is
that at $K=6$ or 8 and $Q=K+8$, the number of equations reduces
from 12 to $P=11$, because of a somewhat simplified structure
of the last truncation operator shown in Eq.~(67) with respect to that
appearing in Eq.~(59). So that below we will present final results only
with brief comments for each the above cases. The best set among the
solutions found were identified as those that minimize the quantity
$\delta=\max_{p=1}^P|d_p|$ (almost always this led to the
minimization of the norm for the main term of truncation errors
as well).

For $K=6$ and $Q=10$ there are five solutions with the best set
\small
\begin{eqnarray}
    d_1 =  \ \  0&.&88480139304442862590773863625720{\rm E}\!+\!00
\nonumber \\ [-2pt]
    d_2 =  \ \  0&.&11922404430206648052593264029266{\rm E}\!+\!00
\\ [-2pt]
    d_3 =      -0&.&10677277516805770678518370004925{\rm E}\!+\!01
\nonumber
\end{eqnarray}
\normalsize
with $d_4=1-2(d_1+d_2+d_3)$ and $\delta_{\rm min} \equiv |d_4| =
1.127$ (within three significant digits).
At $K=8$ and $Q=12$ we have
found again five solutions and the optimal one is
\small
\begin{eqnarray}
    d_1 =  \ \  0&.90803696667238426284572611022928&{\rm E}\!+\!00
\nonumber \\ [-2pt]
    d_2 =  \ \  0&.95777180465215511634906238400062&{\rm E}\!-\!01
\\ [-2pt]
    d_3 =      -0&.10545412798113627599734519738778&{\rm E}\!+\!01
\nonumber
\end{eqnarray}
\normalsize
with $d_4=1-2(d_1+d_2+d_3)$ and $\delta_{\rm min} \equiv |d_4| =
1.101$.

For $K=6$ and $Q=12$ there were more than 150 solutions with the
optimal set
\small
\begin{eqnarray}
    d_1 = \ \ 0&.&64725339206305240605385248392083
\nonumber \\ [-2pt]
    d_2 = \ \ 0&.&44631941526959576960102601257986
\nonumber \\ [-2pt]
    d_3 =    -0&.&66447133641046221008529452937721
\nonumber \\ [-8pt] \\ [-8pt]
    d_4 =    -0&.&58260619571844248816548809046510
\nonumber \\ [-2pt]
    d_5 = \ \ 0&.&64081619589013117205634311707157
\nonumber \\ [-2pt]
    d_6 = \ \ 0&.&31805596598883340430918587031701
\nonumber
\end{eqnarray}
\normalsize
and $d_7=1-2 (d_1+d_2+d_3+d_4+d_5+d_6)$
with $\delta_{\rm min} \equiv |d_3| = 0.664$.

When $K=8$ and $Q=14$ we have computed more than 150 solutions also
and identified among them the following optimal one
\small
\begin{eqnarray}
    d_1 = \ \ 0&.&61158201716899487377123317047417
\nonumber \\ [-2pt]
    d_2 = \ \ 0&.&46763050598682150405078600842681
\nonumber \\ [-2pt]
    d_3 =    -0&.&63245030403272077359889720182431
\nonumber \\ [-8pt] \\ [-8pt]
    d_4 =    -0&.&58223379020720528275072356442667
\nonumber \\ [-2pt]
    d_5 = \ \ 0&.&62109852451075548059651686410928
\nonumber \\ [-2pt]
    d_6 = \ \ 0&.&29686555238409826518407483052733
\nonumber
\end{eqnarray}
\normalsize
with $d_7=1-2 (d_1+d_2+d_3+d_4+d_5+d_6)$
and $\delta_{\rm min} \equiv |d_3| = 0.632$.

At $K=6$ and $Q=14$ the best set, among
more than 200 solutions realized, is

\vspace{9pt}

\small
\begin{eqnarray}
    d_1    = \ \ 0&.&32557163066085080712970217977681
\nonumber \\ [-2pt]
    d_2    =    -0&.&47389771786834222637653653795835
\nonumber \\ [-2pt]
    d_3    = \ \ 0&.&54376649763596364670254533524499
\nonumber \\ [-2pt]
    d_4    =    -0&.&64055411141298491334240825973418
\nonumber \\ [-2pt]
    d_5    = \ \ 0&.&28139025047030322588052971757542
\nonumber \\ [-8pt] \\ [-8pt]
    d_6    = \ \ 0&.&56345778618405675650229011409013
\nonumber \\ [-2pt]
    d_7    = \ \ 0&.&64205004597526944181678051477448
\nonumber \\ [-2pt]
    d_8    =    -0&.&16972825772391310721875128881451
\nonumber \\ [-2pt]
    d_9    =    -0&.&57973031669054683392549871514985
\nonumber \\ [-2pt]
    d_{10} = \ \ 0&.&27398580283063379870623390979762
\nonumber
\end{eqnarray}

\vspace{9pt}

\normalsize
\noindent
with $d_{11}=1-2 \sum_{p=1}^{10} d_p$
and $\delta_{\rm min} \equiv |d_7| = 0.642$.

Finally for $K=8$ and $Q=16$ the optimal solution, among again
more than 200 sets calculated, is

\vspace{9pt}

\small
\begin{eqnarray}
    d_1    = \ \ 0&.&29642254891413070953312450213071
\nonumber \\ [-2pt]
    d_2    = \ \ 0&.&55268563185301488324882994018746
\nonumber \\ [-2pt]
    d_3    =    -0&.&58134339535533393315605544309940
\nonumber \\ [-2pt]
    d_4    = \ \ 0&.&23403665265420481243563202333267
\nonumber \\ [-2pt]
    d_5    =    -0&.&51788958989817055303978658827453
\nonumber \\ [-8pt] \\ [-8pt]
    d_6    =    -0&.&43983975477992920522811970527874
\nonumber \\ [-2pt]
    d_7    =    -0&.&20137078150942169957468111993444
\nonumber \\ [-2pt]
    d_8    = \ \ 0&.&34412872002528894622975927197416
\nonumber \\ [-2pt]
    d_9    = \ \ 0&.&03072591760996558798895428309765
\nonumber \\ [-2pt]
    d_{10} = \ \ 0&.&48652953960727041281280535031455
\nonumber
\end{eqnarray}

\vspace{9pt}

\normalsize
\noindent
with $d_{11}=1-2 \sum_{p=1}^{10} d_p$ and $\delta_{\rm min} \equiv
|d_{11}| = 0.592$.

As can be seen, the quantity $\delta_{\rm min}$ decreases with increasing
$Q$ at any $K$ (4, 6, and 8) considered. Besides the improvement of the
precision of the integration, this leads to an extension of the region
of stability of the composition algorithms. Indeed, the constants $d_p$
appear in the compositions (see Eq.~(54)) in the form of the term $d_p
\Delta t$ (and its combinations of different orders when evaluating
truncation uncertainties ${\cal O}(\Delta t^{Q+1})$). Then, taking into
account that $\delta=\max_{p=1}^P|d_p|$, the maximal value for the size
$\Delta t$ of the time step, at which these uncertainties do not exceed
an acceptable level of precision, can be estimated as ${\Delta t}_{\max}
\sim 1/\delta_{\rm min}$. This also explains a well correlation of
$\delta_{\rm min}$ with the minimum for the norm of truncation errors.

Sixth- and eighth-order based compositions may have advantages
over algorithms basing on fourth-order schemes especially when constructing
very precise integrators with high values of $Q$.
For instance, in order
to derive an integrator of order $Q=16$ on the basis of triplet concatenation
(55) of a scheme of order $K=4$,
it is necessary to apply $3^{(Q-K)/2}=729$ fourth-order stages.
Taking into account that each such a stage requires $n_{\rm f}=3$ force
and $n_{\rm g}=1$ force-gradient evaluations (see Eqs.~(38) or (39)), one
obtains in total the numbers $n_{\rm f}=2187$ and $n_{\rm g}=729$
corresponding to a whole time step. On
the other hand, in view of result (73), an integrator of order $Q=16$
can be composed at $K=8$ and $P=11$ using $2P-1=21$ eighth-order stages
for each of which $n_{\rm f}=n_{\rm g}=11$ (see Eq.~(53)). So
that the overall number of force and force-gradient recalculations will
be equal only to 231 that is much smaller than the above values 2187 and
729 obtained in the case $K=4$.

\section{Applications of force-gradient algorithms}

\subsection{Molecular dynamics simulations}

In molecular dynamics (MD) simulations we dealt with a system of $N=256$
identical ($m \equiv m_i$) particles interacting through a Lennard-Jones
potential, $\Phi(r) = 4 u \big[(\sigma/r)^{12}-(\sigma/r)^6\big]$. The
particles were placed in a cubic box of volume $V=L^3$ and periodic
boundary conditions have been used to exclude the finite-size effects.
For the same reason, the initial potential was modified as $\varphi(r)=
\Phi(r)-\Phi(r_{\rm c})+(r-r_{\rm c}) \Phi'(r_{\rm c})$ at $r \le r_{\rm
c}$ with $\varphi(r)=0$ for $r > r_{\rm c}$, where $r_{\rm c}=L/2$ is the
cut-off radius. Then the potential $\varphi$ and its first-order
derivative $\varphi'={\rm d} \varphi /{\rm d} r$ will be continuous
functions any where in $r$ including the truncation point $r=r_{\rm c}$.
This avoids an energy drift caused by the passage of particles
via the surface of truncation sphere
as well as singularities of $\varphi'(r)$
and $\varphi''(r)$ at $r=r_{\rm c}$. The simulations were carried out
in a microcanonical ensemble at a reduced density of $n^\ast=N \sigma^3/
V=0.845$ and a reduced temperature of $T^\ast=k_{\rm B}T/u=1.7$.
All runs of the length in $l=10\,000$ time steps each were started from
an identical well equilibrated initial configuration ${\mbox{\boldmath
$\rho$}}(0)$. The precision of the integration was measured in terms of
the relative total energy fluctuations ${\cal E}=\langle (E- \langle E
\rangle)^2 \rangle/|\langle E \rangle|$, where $E=\frac12 \sum_{i=1}^N
m {{\bf v}_i}^2+\frac12 \sum_{i \ne j}^N \varphi(r_{ij})$ and
$\langle \ \rangle$ denotes the microcanonical averaging.

The equations of motion were integrated using force-gradient algorithms
(30), (36), and (38) of the fourth order within schemes A, A$'$, B, C,
and C$'$ (see Eqs. (32), (33), (37), (41), and (42), respectively). For the
purpose of comparison the integration with the help of usual fourth-order
algorithm by Forest and Ruth (FR) \cite{Forest} (which represents, in fact,
triplet concatenation (55) of second-order Verlet scheme (8)) has been
performed as well. The corresponding results for the total energy
fluctuations as functions of the length $l=t/\Delta t$ of the simulations
is presented in subset (a) of Fig.~1 for a typical reduced time step of
$\Delta t^\ast=\Delta t (u/m \sigma^2)^{1/2}=0.005$. As can be seen,
schemes A, B, and C exhibit a similar equivalence in energy conservation.
This is in agreement with our theoretical predictions presented in
subsection
II.B.{\em 4}, where the precision of algorithms has been estimated
in terms of the norm $\gamma$ (Eq.~(34)) of fifth-order truncation
errors ${\cal O}(\Delta t^5)$. In particular for schemes A, B, and C,
it has been obtained $\gamma \approx 0.000713$, 0.000715, and 0.000715,
respectively. Further, as expected, scheme A$'$ ($\gamma
\approx 0.00334$) is worse in precision and leads to values of ${\cal E}$
which are approximately in $0.00334/0.00071 \approx 5$ times larger.
Note that in microcanonical ensembles the total energy is an integral
of motion, $E(t) = E(0)$, so that within approximate MD simulations,
smaller values of ${\cal E}$ correspond to a better precision of the
integration. It is worth remarking also that another integral of motion,
namely, total momentum ${\bf P}=\sum_{i=1}^N m_i {\bf v}_i$,
is conserved exactly within force-gradient approach (13). The reasons are
that all velocities are updated at once (see Eq.~14) during each stage of
decompositions and the fact that $\sum_{i=1}^N {\bf f}_i=0$ as well
as $\sum_{i=1}^N {\bf g}_i=0$ (as can be verified readily using the
structure of Eq.~(11)).

The best accuracy in energy conservation can be achieved within optimized
scheme C$'$ (see Fig.~1 (a)) for which $\gamma \approx 0.000141$. It
minimizes ${\cal E}$ to a level of $\sim 10^{-5}$ that is in factor
$0.00071/0.000141 \approx 5$ lower than those related to schemes A, B,
and C. At the same time, the usual FR algorithm leads to the worst result
${\cal E} \sim 10^{-3}$. We see, therefore, that applying force-gradient
approach allows to reduce unphysical energy fluctuations up in two orders
in magnitude. Let us show now that this overcompensates an increased
computational efforts caused by additional calculations of the force
gradients. The processor time used for carrying out these calculations
(see Eq.~(11)) was nearly in 3 times larger than that needed for evaluations
of forces itself. Further, we should take into account that algorithm C$'$
requires $n_{\rm f}=3$ force and $n_{\rm g}=1$ force-gradient recalculations
per time step, whereas $n_{\rm f}=3$ and $n_{\rm g}=0$ for FR scheme. As
a result, one obtains that the size $\Delta t$ of one step within FR
propagation must be in $(3 + 3 \cdot 1)/3=2$ times shorter than in
the case of algorithm C$'$ for spending the same overall processor
time within both the cases during the integration over a fixed time
interval. Finally, in view of the fact that the global error and thus the
function ${\cal E}$ are proportional to the fourth power of $\Delta t$,
i.e., ${\cal E} \sim \Delta t^4$, one finds that, at the above conditions,
the level of conservation of the FR scheme reduces from $10^{-3}$ to
${\cal E} \sim 10^{-3}/2^4$. So that relative efficiency of scheme C$'$
with respect to the FR integrator is actually equal to $(10^{-3}/2^4)/
10^{-5}=100/16 \approx 6$.

In order to ensure that scheme C$'$ (Eq.~(42)) is indeed the best among
whole family (40) of C$'$-like integrators (38), we carried out additional
simulations in which the parameter $\lambda$, being constant within each
simulation, varied from one run to another. The total energy fluctuations
obtained in such simulations at the end of the runs for two (fixed within
each run) undimensional time steps, namely, $\Delta t^\ast=0.0025$
and 0.005,
are shown in subset (b) of Fig.~1 as functions of $\lambda$. As can be
observed, the dependencies ${\cal E}(\lambda,\Delta t)$ have the global
minimum located at the same point $\lambda \approx 0.247$ independently on
the size $\Delta t$ of the time step. This point coincides completely with
the minimum given by Eq.~(42) for the function $\gamma(\lambda)$ (see
Eq.~(34) and relations following just after Eq.~(41)) which is also
included in the subset and plotted by a dashed curve (where an upper
lying part of the curve corresponds to sign plus in Eq.~(40),
whereas a lower lying part as well as the simulation data
are related to sign minus). So that our criterion on measuring the
precision of the integration in terms of the norm
of local truncated
uncertainties is in excellent accord.
Moreover, the energy fluctuations appear to
be proportional to that norm $\gamma$ as
${\cal E}(\lambda,\Delta t) \sim \gamma \Delta t^4$ and the coefficient
of the resulting proportionality almost does not depend on $\lambda$
and $\Delta t$.

\end{multicols}

\vspace{2pt}
\begin{figure}[htbp]
\epsfxsize=160mm \centerline{\epsffile{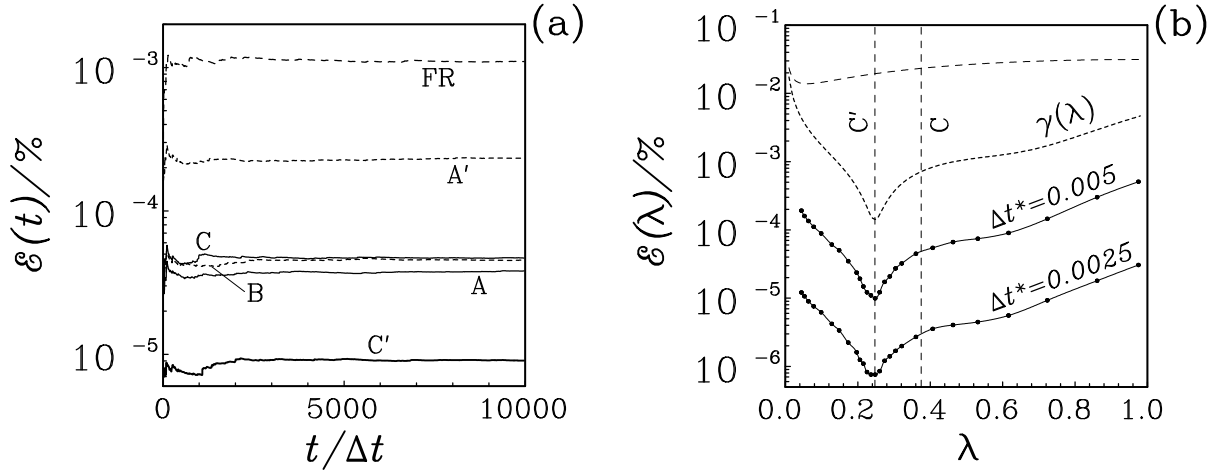}} \vspace{2mm}
\caption{(a) The total energy fluctuations ${\cal E}$ as functions
of the length of the simulations performed using force-gradient
algorithms A, A$'$, B, C, and C$'$ in comparison with the result
of usual FR scheme. (b) The fluctuations ${\cal E}$ obtained
within extended C$'$-like scheme as depending on free parameter
$\lambda$ at two fixed time steps. The values of $\lambda$ related
to schemes C and C$'$ are marked by the vertical lines. The
function $\gamma (\lambda)$ is plotted by the dashed curve (see
the text for additional explanations).}
\end{figure}

\vspace{14pt}

\begin{multicols}{2}

It is worth remarking that the results reported in this subsection should
be considered as the first attempts of applying force-gradient algorithms
to MD simulations. In previous studies, algorithms of such a kind have been
tested for classical \cite{Chin,Chins} and quantum \cite{ChinChen} mechanics
systems composed of a few bodies only (or even one body moving in an external
field). The present investigations have demonstrated that force-gradient
algorithms can be used with equal success in statistical mechanics
simulations dealing with a great number of particles, i.e., when $N \gg 1$.
In the last case, the calculations of force gradients also
presents no difficulties. Indeed, during the integration we should first
evaluate usual forces ${\bf f}_i$ for each particle $i$, where $i=1,2,
\ldots,N$. This involves a number of operations which is proportional
to the second power of $N$.
Then in view of the structure of Eq.~(11)
and taking into account the fact that particle's accelerations ${\bf a}_i=
{\bf f}_i/m_i$ are already known quantities, the calculations of gradients
${\bf g}_i$ will require a number of operations which is proportional to
the same power of $N$, i.e., $\propto N^2$
(but not to $\propto N^3$, as it may look at first sight).
Further reducing
the computational efforts is possible taking
into account that function ${\bf g}({\bf r}_{ij})$ (see Eq.~(11))
decreases with
increasing the interparticle separation $r_{ij}$ more rapidly than
initial potential $\varphi(r_{ij})$. In such a case, a secondary cut-off
radius $R_{\rm c} < r_{\rm c}$ can be introduced when evaluating ${\bf
g}_i$ (that is equivalent to putting ${\bf g}({\bf r}_{ij}) = 0$ at
$r_{ij} > R_{\rm c}$) in order to speed up the calculations.

\subsection{Celestial mechanics simulations}

One of the simplest way to test force-gradient algorithms of higher orders
is to apply them to solution of the two-dimensional Kepler problem. In
particular, this way has been chosen by Chin and Kidwell \cite{Chin,Chins}
when testing fourth-order algorithms A, B, and C and higher-order iterated
counterparts of the last scheme. As has been established, this force-gradient
scheme is particularly outstanding and appears to be much more superior than
usual non-gradient integrators, such as fourth-order by Forest and Ruth
\cite{Forest} as well as by Runge and Kutta \cite{Kinos,Gladm}, sixth-order
by Yoshida \cite{Yoshida}, etc. In this subsection it will demonstrated that
further significant improvement in the effectiveness of the integration
can be reached replacing standard iteration procedure (55) by advanced
composition approach (54). Moreover, using our new sixth- and eighth-order
force-gradient algorithms as the basis for the composition has allowed us
to perform the computations with extremely high precision which exceeds by
several orders the accuracy observing within standard fourth-order based
schemes.

We will consider a motion of a particle (planet) of mass $m_1$
moving in the (gravitation) field $\varphi(r)=-c/r$ of the central body
(sun) with mass $m_2 \gg m_1$, where $c>0$ is the constant responsible
for intensity of the interaction. For simplifying the calculations, one
neglects the influence of all other ($i=3,4,\ldots,N$) particles (planets,
for which $m_i \ll m_2$) in the (solar) system. Then the motion can be
described by the following system of two equations,
\begin{equation}
\frac{{\rm d} {\bf r}}{{\rm d} t} = {\bf v} \, ,
\ \ \ \ \ \ \
\frac{{\rm d} {\bf v}}{{\rm d} t} = - \frac{\bf r}{r^3} \, ,
\end{equation}

\vspace{2pt}

\noindent
where ${\bf r}={\bf r}_1-{\bf r}_2$, and
for clarity of the presentation we have used units in which the
reduced mass $m_1 m_2 /(m_1+m_2)$ and the interaction constant $c$ are equal
to unity. Since the quantity $E=v^2/2-1/r$ (which is associated with the
total energy) presents an integral of motion for equations (74), it should be
conserved during the integration. However, this will so if these equations
are solved exactly. In numerical simulations, the local truncation
uncertainties ${\cal O}(\Delta t^{Q+1})$ accumulate step by step of the
integration process, leading at $t \gg \Delta t$ to the global errors
${\cal O}(\Delta t^Q)$, where $Q$ denotes the order of a self-adjoint
algorithm. So that the quantity $E$ can be presented as a function of time
as
\begin{equation}
E(t) = E_0 + E_Q(t) \Delta t^Q + {\cal O}(\Delta t^{Q+2}) \, ,
\end{equation}
where $E_0 \equiv E(0)$ and $E_Q$ is the main step-size independent error
coefficient.

In our simulations we solved two-dimensional Kepler problem (74) with
the same initial conditions ${\bf r}(0)=(10,0)$ and ${\bf v}(0)=(0,1/10)$
as those used by previous authors \cite{Chin,Chins} to make comparative
analysis more convenient.
The resulting highly eccentric ($e=0.9$) orbit provides
a nontrivial testing ground for trajectory integration.
The numerical effectiveness of each algorithm was
gauged in terms of main error coefficient $E_Q=\lim_{\Delta t \to 0}
[E(t)-E_0]/\Delta t^Q$ (see Eq.~(75)).
It
can actually be
extracted from the fraction $[E(t)-E_0]/\Delta t^Q$ by choosing smaller
and smaller time steps $\Delta t$ to be entitled to completely
ignore next
higher-order corrections ${\cal O}(\Delta t^{Q+2})$.
This typically occurs in the neighborhood of $\Delta t \sim P/5000$, where
$P = \pi /(2 |E_0|^3)^{1/2}$ is the period of the elliptical orbit.
Since we are
dealing with algorithms of high orders $Q$ and small step sizes
$\Delta t$, all the calculations have been carried out in Fortran using
quadruple (instead of double, as for MD simulations) precision
arithmetics for
ensuring
the correctness of the results.

The normalized energy deviations $E_Q/E_0$ obtained in the simulations
applying fourth-, sixth-, eighth-, tenth-, twelfth-, and fourteenth-order
algorithms are plotted in subsets (a), (b), (c), (d), (e), and (f) of
Fig.~2, respectively, as functions of time $t$ during one period $P$ of the
orbit. These deviations are substantial only near mid period when the
particle is at its closest position to the attractive center. Note also
that within symplectic integration, the nonconservation
of energy for periodic orbits is periodic and its averaged (over times
$t \gg P$) value is bounded and independent on $t$ (the independence
of averaged energy fluctuations at $t \gg \Delta t$ has already been
demonstrated in MD simulations, see Fig.~1). That is way we presented
the results in Fig.~2 within a narrow region of time near $t \sim P/2$,
where the maximal deviations of $E_Q$ will give a main contribution to
the overall fluctuations.

In the case of fourth-order integration
we used most typical algorithms
A, B, C, and C$'$ (see Eqs. (32), (37), (41), and (42), respectively).
As can be seen from subset (a) of Fig.~2, the pattern here is somewhat
different than that in MD simulations (please compare with Fig.~1 (a)).
The algorithm C is clearly better than schemes A and B, that
confirms the conclusion of Ref.~\cite{Chin}. On the other hand,
integrator C$'$ does not exhibit an improved precision in
energy conservation with respect to scheme C. Nearly the same was seen
when iterating these algorithms to higher orders
with the help of triplet construction (55).
In particular, the sixth-order C$'$-counterpart appeared
to be even slightly better than the corresponding counterpart of scheme
C (see subset (b) of Fig.~2). At the same time, higher-order integrators
basing on schemes A and B were definitely worse. So that the obvious
candidates for fourth-order based iterations (55) and
compositions (54) are schemes C and C$'$.

In order to understand why scheme C$'$ does not lead to the expected
improvement over scheme C in this particular situation, it should be
taken into account
that we deal with a small system, actually with one body moving in
an effective external field. Moreover, such a body moves periodically and,
thus, covers only small part of phase space during its displacement.
This is
contrary to many-body statistical systems, where the phase point may
visit considerably wider regions of phase space. In the latter case,
during the averaging along the phase trajectories, different components
$\gamma_{1-4}$ of fifth-order local uncertainties
(see Eq.~(17)) will enter with approximately the same weights when forming
the total error vector
${\cal O}(\Delta t^5)$.
This has been tentatively assumed when writing the
norm $\gamma$ of that vector in the form of Eq.~(34) and further minimizing
$\gamma$
to obtain algorithm C$'$. In the case of a few-body system,
especially with periodic motion, the above weights may differ considerably.
This complicates an analysis of the truncation terms and makes impossible
to find an exact global minimum for them within any analytical approach.
Note, however, that even here, the assumption on uniform contribution of
truncation-error components works relatively well. Indeed, in view of
dependencies shown in subsets (a) and (b), we can stay that both the schemes
C and C$'$ are comparable in precision. The same was observed for their
higher-order counterparts. For this reason (and to reserve more free space
for other dependencies), in further subsets (c)--(f) we will draw only
curves corresponding to scheme C.

\end{multicols}

\begin{figure}[htbp]
\epsfxsize=160mm \centerline{\epsffile{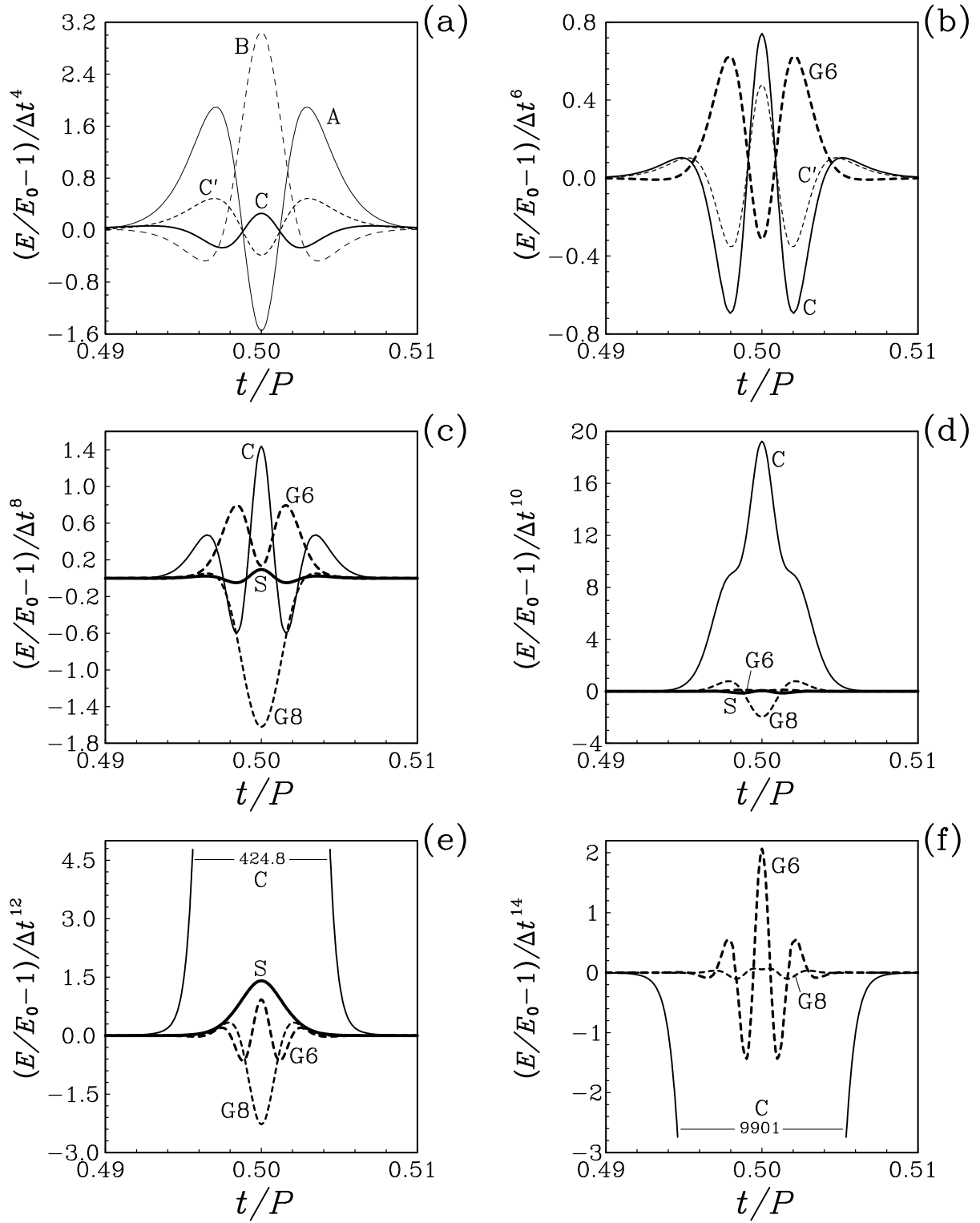}} \vspace{2mm}
\caption{The normalized energy deviation of a particle in a
Keplerian orbit. The results obtained within fourth-, sixth-,
eighth-, tenth-, twelfth-, and fourteenth-order algorithms are
shown in subsets (a), (b), (c), (d), (e), and (f), respectively.
The basic algorithms used are: fourth-order schemes A, B, C, and
C$'$, as well as sixth- and eight-order integrators
(correspondingly marked as G6 and G8). The curves related to
higher-order algorithms concatenated on the basis of schemes C by
standard iterations are labelled by the same letter C in each the
sets. The fourth-, sixth-, and eighth-order based algorithms
constructed within advanced composition approach are marked as S,
G6 and G8, respectively (see the text).}
\end{figure}

\begin{multicols}{2}

When considering the sixth-order integration, we realized that direct
velocity-like scheme defined by Eqs.~(45) and (47) is much worse (the maximum
deviation of $E_Q$ were more than in two orders larger) than its extended
position-like counterpart given by Eqs.~(50) and (51). This is in agreement
with a prediction of subsection II.B.{\em 5}. The result corresponding to
the position-like algorithm is plotted in subset (b) of Fig.~1 by the
bold dashed curve marked as G6. As can be seen, all three curves shown
in this subset,
namely C, C$'$, and G6 are close enough to each other. But algorithm G6
uses only
$n_{\rm f}=5$ force evaluations per time step, instead of $n_{\rm f}=9$
needed for iterated C- and C$'$-like schemes (for all these three cases
the number of force-gradient evaluations is the same
and equal to $n_{\rm g}=3$).
Therefore, for order six,
direct decomposition approach (13) leads to more efficient integration
than concatenations of fourth-order schemes.

Beginning from order eight, the above concatenations based on standard
iterations (55) and advanced compositions (54) will result in completely
different integrators. The simulation data for
these iterated and composed C-based integrators
are shown in subset (c) of Fig.~2 by thin (marked simply as C) and
bold (marked as S) solid curves, respectively. The curves related to
tenth- and twelfth-order iteration and composition integrators (based
on the same fourth-order scheme C) are plotted correspondingly in
subsets (d) and (e) of Fig.~2,
and marked by the same letters C and S. We mention
that C-marked curves have already been presented in the work by Chin
and Kidwell \cite{Chin,Chins} up to order 12. They are redisplayed by us
in order to illustrate the evident superiority of our new composition
approach over the standard iteration method. Indeed,
for the iteration integrators (C-marked curves) of orders $Q=8$, 10, and
12, the magnitudes of the normalized energy coefficient $E_Q/E_0$ after
one period are 1.44, 19.24, and 424.8, respectively. On the other hand,
the magnitudes related to the composition integrators (S-marked curves)
constitute correspondingly 0.0953, 0.0577, and 1.41, i.e., they are
approximately in 15, 330, and 300 times smaller. In addition, the
composition integrators are faster with respect to their iteration versions
in factors 9/7, 27/13, and 81/23 for $Q=8$, 10, and 12, respectively
(see subsection III.A.{\em 1}), and thus the resulting efficiencies
will increase yet.

What about sixth- and eight-order-based composition schemes at $Q \ge
8$? First of all, let us consider the case of eight-order integration.
Here, the direct scheme chosen was position-like integrator (53) (it
leads to better energy conservation with respect to its velocity-like
counterpart (52)). The result corresponding to this integrator is plotted
in subset (c) of Fig.~2 by the dashed curve marked as G8. As can be seen,
the fourth-order-based composition scheme (S-curve) is better at $Q=8$
with respect to both direct G8 and iterated G6-like versions. With
increasing the order to 10 and 12, all they become to be nearly equivalent
in the accuracy of energy conservation. But fourth-order-based approach
requires somewhat fewer number of operations. For instance, for order 12,
one obtains that the numbers of force and gradients evaluations per time
step are equal for it to $n_{\rm f}=23 \cdot 3=69$ and $n_{\rm g}=23$,
respectively, whereas
these numbers for sixth- and
eighth-order-based compositions G6 and G8 are $n_{\rm f}=13 \cdot 5=65$,
$n_{\rm g}=13 \cdot 3=39$ and $n_{\rm f}=n_{\rm g}=
7 \cdot 11=77$
(where G6-integrator requires less operations than G8-scheme).
However, beginning from order 14, the situation reverses. The
fourth-order-based composition S-approach becomes to be not longer
accessible (because of the absence of explicit expressions for its
time coefficients here). On the other hand, applying the standard
fourth-order-based iteration C-method is very inefficient.
In particular, at $Q=14$ the maximal energy deviations within this method
is $|E_{14}/E_0|_{\rm max}=9901$ with $n_{\rm f}=21 \cdot 5=729$
and $n_{\rm g}=21 \cdot 3=243$. At the same time, the higher-order-based
composition schemes lead to much accurate
results, namely, $|E_{14}/E_0|_{\rm
max}=2.065$ with $n_{\rm f}=21 \cdot 5=105$ and $n_{\rm g}=21 \cdot 3=63$
for G6- as well as $|E_{14}/E_0|_{\rm max}=0.101$
with $n_{\rm f}=n_{\rm g}=13 \cdot
11=143$ for G8-based schemes (where the
better precision for the last scheme compensates to some extent its
increased values for quantities $n_{\rm f}$ and $n_{\rm g}$). We see,
therefore, that the relative efficiencies of G6- and G8-based schemes
with respect to C-approach constitute about $10^4$--$10^5$. Finally,
in the case $Q=16$ (not shown in Fig.~2) we have obtained the values
$|E_{16}/E_0|_{\rm max}=2.43 \cdot 10^5$ and $|E_{16}/E_0|_{\rm max}=
48.16$ corresponding to C- and G8-based schemes, respectively. Taking
into account the numbers of $n_{\rm f}$ and $n_{\rm g}$ for these
schemes presented at the end of subsection II.C.{\em 2}, one can
conclude that the efficiency increases here also approximately in
$10^4$ to $10^5$ times.

\section{Concluding remarks}

In this work we have formulated a general theory of construction
of force-gradient algorithms for solving the equations of motion
in classical and quantum systems. This has allowed us to extend
considerably the class of analytically integrable symplectic
schemes. The new algorithms derived include self-adjoint direct
decomposition integrators of orders two, four, six, and eight
as well as their composition counterparts up to the sixteenth
order in the time step. As has been proven theoretically and
confirmed in actual numerical simulations, these algorithms lead
to significant improvement in the efficiency of the integration
with respect to existing force-gradient and non-gradient schemes.
It has been demonstrated that force-gradient algorithms can be
used with equal success as for describing the motion in few-body
classical and quantum mechanics systems as well as for performing
statistical molecular dynamics observations over many-particle
collections. In all the cases the calculation of force gradients
presents no difficulties and requires computational efforts
comparable with those needed to evaluate usual forces itself.
The new algorithms may be especially useful for the prediction
and study of very subtle dynamical effects in different areas of
physics and chemistry including the problems of astrophysical
interest, whenever the precise integration of motion during
very long times is desirable.

The algorithms introduced exactly reproduce such important features
of classical dynamics as time reversibility and symplecticity.
This explains their excellent energy conservation and stability
properties. In this context it should be mentioned another class of
(non-gradient) integrators recently developed \cite{Sofron} on the
basis of a modified Runge-Kutta approach. Like the force-gradient
algorithms, the Runge-Kutta-like integrators also allow to produce
time-reversible and symplectic trajectories in phase space with,
in principle, arbitrary order in precision. However, such integrators
are implicit and require cumbersome systems of globally coupled (via
positions and forces of all particles) nonlinear equations be solved
by expensive iterations at each step of the integration process.
Since in practice such equations cannot be solved exactly, the time
reversibility and symplecticity can be violated. This may lead, in
particular, to instabilities in long-term energy conservation,
i.e., to the same problem inherent in the tradition (nonsymplectic)
Runge-Kutta method (see the Introduction). All these disadvantages
are absent in the present approach, where the phase trajectories are
propagated explicitly in time by applying consecutive simple shifts
of particles in velocity and position space with exact preservation
of the phase volume and reversibility of the generated solutions.

The approach presented can also be adapted to the integration of motion
in more complicated systems, such as systems with orientational or spin
degrees of freedom, etc., where splitting of the Liouville operator into
more than two parts may be necessary to obtain analytically solvable
subpropagators. These and other related problems will be considered
in a separate investigation.

\section*{Acknowledgment}

Part of this work was supported by the Fonds zur F\"orderung der
wissenschaftlichen Forschung under Project No.~15247.

\end{multicols}

\vspace{12pt}

\begin{center}
{\bf
Appendix}
\end{center}
\setcounter{equation}{0}
\renewcommand{\theequation}{A\arabic{equation}}

\begin{multicols}{2}

The recursive relations for the highest-order multipliers $\zeta_{1-10}$
(see Eqs.~(13), (15), and (18)) corresponding to the first type of
self-adjoint transformations given by the first line of Eq.~19 are:

\end{multicols}

\vspace{6pt}

\begin{eqnarray*}
\zeta_1^{(n+1)} =\zeta_1^{(n)} +
&&
     a^{(n)} \big(630 {\beta^{(n)}}^2 + 1260 \gamma_4^{(n)} \sigma^{(n)} -
63 \beta^{(n)} \big(6 a^{(n)} + \nu^{(n)}\big) {\sigma^{(n)}}^2 +
\\
&&
   {\sigma^{(n)}}^3 \big(21 \alpha^{(n)} + \big(27 ({a^{(n)}}^2 +
9 a^{(n)} \nu^{(n)} + {\nu^{(n)}}^2\big) \sigma^{(n)}\big)\big)/3780
\, ,
\end{eqnarray*}
\begin{eqnarray*}
\zeta_2^{(n+1)} =\zeta_2^{(n)} +
&&
     a^{(n)} \big(
336 \beta^{(n)} \big(6 a^{(n)} + \nu^{(n)}\big) {\sigma^{(n)}}^2
-5040 {\beta^{(n)}}^2 - 5040 \gamma_4^{(n)} \sigma^{(n)}
-
\\
&&
   {\sigma^{(n)}}^3 \big( 336 \alpha^{(n)} + \big( 120 ({a^{(n)}}^2 +
12 a^{(n)} \nu^{(n)} - {\nu^{(n)}}^2\big) \sigma^{(n)}\big)\big)/45360
\, ,
\end{eqnarray*}
\begin{eqnarray*}
\zeta_3^{(n+1)} =\zeta_3^{(n)}
-
&&
    a^{(n)} \big(5040 \alpha^{(n)} \beta^{(n)} +
\sigma^{(n)} \big(5040 \gamma_2^{(n)} - 84 \beta^{(n)} {\nu^{(n)}}^2 +
      72 ({a^{(n)}}^3 {\sigma^{(n)}}^2 + {\nu^{(n)}}^3 {\sigma^{(n)}}^2 +
\\
&&
      24 a^{(n)} \nu^{(n)}
\big( \nu^{(n)} {\sigma^{(n)}}^2 -42 \beta^{(n)}
\big) +
      ({a^{(n)}}^2 \big(
88 \nu^{(n)} {\sigma^{(n)}}^2
-672 \beta^{(n)}
\big)
\big)
\big)/15120
\, ,
\end{eqnarray*}
\begin{eqnarray*}
\zeta_4^{(n+1)} =\zeta_4^{(n)} +
&&
     a^{(n)} \big(168 \alpha^{(n)} \big(60 \beta^{(n)} -
\big(6 a^{(n)} + \nu^{(n)}\big) {\sigma^{(n)}}^2\big) +
   \sigma^{(n)} \big(10080 \gamma_2^{(n)} + 5040 \gamma_3^{(n)} -
168 \beta^{(n)} {\nu^{(n)}}^2 +
\\
&&
     192 ({a^{(n)}}^3 {\sigma^{(n)}}^2 +
5 {\nu^{(n)}}^3 {\sigma^{(n)}}^2 +
     6 a^{(n)} \nu^{(n)}
\big(13 \nu^{(n)} {\sigma^{(n)}}^2-336 \beta^{(n)}\big) +
\\
&&
 ({a^{(n)}}^2 \big(272 \nu^{(n)} {\sigma^{(n)}}^2
 -1344 \beta^{(n)}
\big)\big)\big)/15120
\, ,
\end{eqnarray*}
\begin{eqnarray*}
\zeta_5^{(n+1)} =\zeta_5^{(n)}
-
&&
    a^{(n)} \big(2520 \gamma_4^{(n)} \nu^{(n)} +
7560 \gamma_3^{(n)} \sigma^{(n)} - 294 \beta^{(n)} {\nu^{(n)}}^2 \sigma^{(n)} +
    180 ({a^{(n)}}^3 {\sigma^{(n)}}^3 - {\nu^{(n)}}^3 {\sigma^{(n)}}^3 +
    \\
&&
    84 \alpha^{(n)} \big(120 \beta^{(n)} + \big(
3 \nu^{(n)}
-22 a^{(n)}
\big) {\sigma^{(n)}}^2\big) +
    ({a^{(n)}}^2 \big(
234 \nu^{(n)} {\sigma^{(n)}}^3
-1512 \beta^{(n)} \sigma^{(n)}
\big) +
\\
&&
    6 a^{(n)} \big(420 \gamma_4^{(n)} -
308 \beta^{(n)} \nu^{(n)} \sigma^{(n)} +
3 {\nu^{(n)}}^2 {\sigma^{(n)}}^3\big)\big)/45360
\, ,
\\ [-30pt]
\end{eqnarray*}
\begin{equation}
\end{equation}
\begin{eqnarray*}
\\ [-19pt]
\zeta_6^{(n+1)} =\zeta_6^{(n)} +
&&
     a^{(n)} \big(18 ({a^{(n)}}^3 {\sigma^{(n)}}^3 -
84 \alpha^{(n)} \big(15 \beta^{(n)} - \big(a^{(n)} +
\nu^{(n)}\big) {\sigma^{(n)}}^2\big) +
   ({a^{(n)}}^2 \big(
15 \nu^{(n)} {\sigma^{(n)}}^3
-
252 \beta^{(n)} \sigma^{(n)}
\big) +
\\
&&
   6 a^{(n)} \big(210 \gamma_4^{(n)} -
28 \beta^{(n)} \nu^{(n)} \sigma^{(n)} +
{\nu^{(n)}}^2 {\sigma^{(n)}}^3\big) +
   2 \big(630 \gamma_4^{(n)} \nu^{(n)} -
\\
&&
630 \gamma_2^{(n)} \sigma^{(n)} -
42 \beta^{(n)} {\nu^{(n)}}^2 \sigma^{(n)} +
     {\nu^{(n)}}^3 {\sigma^{(n)}}^3\big)\big)/7560
\, ,
\end{eqnarray*}
\begin{eqnarray*}
\zeta_7^{(n+1)} =\zeta_7^{(n)} +
&&
     a^{(n)} \big(2520 {\alpha^{(n)}}^2 - 84 \alpha^{(n)} \big(8 ({a^{(n)}}^2 +
12 a^{(n)} \nu^{(n)} + {\nu^{(n)}}^2\big) \sigma^{(n)} +
   \sigma^{(n)} \big(5040 \gamma_1^{(n)} +
\\
&&
\big(48 ({a^{(n)}}^4 +
120 ({a^{(n)}}^3 \nu^{(n)} + 92 ({a^{(n)}}^2 {\nu^{(n)}}^2 +
       18 a^{(n)} {\nu^{(n)}}^3 + {\nu^{(n)}}^4\big) \sigma^{(n)}\big)\big)/15120
\, ,
\end{eqnarray*}
\begin{eqnarray*}
\zeta_8^{(n+1)} =\zeta_8^{(n)}
-
&&
    a^{(n)} \big(5040 {\alpha^{(n)}}^2 + 2520 \gamma_2^{(n)} \nu^{(n)} -
42 \beta^{(n)} {\nu^{(n)}}^3 +
    2520 \gamma_1^{(n)} \sigma^{(n)} -
420 \alpha^{(n)} a^{(n)} \big(a^{(n)} + 2 \nu^{(n)}\big) \sigma^{(n)} +
\\
&&
    69 ({a^{(n)}}^4 {\sigma^{(n)}}^2 + {\nu^{(n)}}^4 {\sigma^{(n)}}^2 +
 2 ({a^{(n)}}^2 \nu^{(n)} \big(
53 \nu^{(n)} {\sigma^{(n)}}^2
-294 \beta^{(n)}
\big) +
    ({a^{(n)}}^3 \big(
148 \nu^{(n)} {\sigma^{(n)}}^2
-294 \beta^{(n)}
\big) +
\\
&&
    6 a^{(n)} \big(420 \gamma_2^{(n)} - 56 \beta^{(n)} {\nu^{(n)}}^2 +
3 {\nu^{(n)}}^3 {\sigma^{(n)}}^2\big)\big)/15120
\, ,
\end{eqnarray*}
\begin{eqnarray*}
\zeta_9^{(n+1)} =\zeta_9^{(n)} +
&&
     a^{(n)} \big(2520 {\alpha^{(n)}}^2 - 42 \alpha^{(n)} \big(8 ({a^{(n)}}^2 +
12 a^{(n)} \nu^{(n)} + {\nu^{(n)}}^2\big) \sigma^{(n)} +
   114 ({a^{(n)}}^4 {\sigma^{(n)}}^2 - 4 ({a^{(n)}}^3 \big(147 \beta^{(n)} -
   \\
&&
59 \nu^{(n)} {\sigma^{(n)}}^2\big) +
   ({a^{(n)}}^2 \nu^{(n)} \big(
173 \nu^{(n)} {\sigma^{(n)}}^2
-1176 \beta^{(n)}
\big) +
   24 a^{(n)} \big(210 \gamma_2^{(n)} + 105 \gamma_3^{(n)} -
28 \beta^{(n)} {\nu^{(n)}}^2 +
   \\
&&
2 {\nu^{(n)}}^3 {\sigma^{(n)}}^2\big) +
   \nu^{(n)} \big(5040 \gamma_2^{(n)} + 2520 \gamma_3^{(n)} -
84 \beta^{(n)} {\nu^{(n)}}^2 + 5 {\nu^{(n)}}^3 {\sigma^{(n)}}^2\big)\big)/
 15120
\, ,
\end{eqnarray*}
\begin{eqnarray*}
\zeta_{10}^{(n+1)} =\zeta_{10}^{(n)} +
&&
     a^{(n)} \big(a^{(n)} + \nu^{(n)}\big) \big(2520 \gamma_1^{(n)} -
42 \alpha^{(n)} \big(7 ({a^{(n)}}^2 + 7 a^{(n)} \nu^{(n)} + {\nu^{(n)}}^2\big) +
\\
&&
   \big(31 ({a^{(n)}}^4 + 62 ({a^{(n)}}^3 \nu^{(n)} +
42 ({a^{(n)}}^2 {\nu^{(n)}}^2 + 11 a^{(n)} {\nu^{(n)}}^3 +
{\nu^{(n)}}^4\big) \sigma^{(n)}\big)/
 15120
\, .
\end{eqnarray*}

\vspace{12pt}

For the transformation of the second type (see the second line of Eq.~(19))
we have obtained:

\vspace{9pt}

\begin{eqnarray*}
\zeta_1^{(n+1)} =\zeta_1^{(n)} -
&&
\big( 18 {b^{(n)}}^4 {\nu^{(n)}}^3 + 15 {b^{(n)}}^3 {\nu^{(n)}}^3 \sigma^{(n)}
+ 42 c^{(n)} \nu^{(n)} \big(30 \beta^{(n)} +
30 c^{(n)} - \nu^{(n)} {\sigma^{(n)}}^2\big) -
\\
&&
  84 \alpha^{(n)} \big(15 \beta^{(n)} b^{(n)} +
30 b^{(n)} c^{(n)} - 3 {b^{(n)}}^3 \nu^{(n)} + 15 c^{(n)} \sigma^{(n)} -
    2 {b^{(n)}}^2 \nu^{(n)} \sigma^{(n)} -
b^{(n)} \nu^{(n)} {\sigma^{(n)}}^2\big) -
\\
&&
  6 {b^{(n)}}^2 \big(210 \gamma_2^{(n)} + {\nu^{(n)}}^2 \big(14 \beta^{(n)}
+
63 c^{(n)} - \nu^{(n)} {\sigma^{(n)}}^2\big)\big) +
  b^{(n)} \big( 1260 \gamma_4^{(n)} \nu^{(n)} -
\\
&&
2 \sigma^{(n)} \big(630 \gamma_2^{(n)} +
      {\nu^{(n)}}^2 \big(42 \beta^{(n)} + 84 c^{(n)} -
\nu^{(n)} {\sigma^{(n)}}^2\big)\big)\big)\big)/7560
\, ,
\end{eqnarray*}
\begin{eqnarray*}
\zeta_2^{(n+1)} =\zeta_2^{(n)} +
&&
\big(12 {b^{(n)}}^4 {\nu^{(n)}}^3 - 39 {b^{(n)}}^3 {\nu^{(n)}}^3 \sigma^{(n)} +
  42 c^{(n)} \nu^{(n)} \big(120 \beta^{(n)} +
120 c^{(n)} - \nu^{(n)} {\sigma^{(n)}}^2\big) -
  252 \alpha^{(n)} \big(20 \beta^{(n)} b^{(n)} +
\\
&&
40 b^{(n)} c^{(n)} - 3 {b^{(n)}}^3 \nu^{(n)} + 20 c^{(n)} \sigma^{(n)} -
    2 {b^{(n)}}^2 \nu^{(n)} \sigma^{(n)} -
b^{(n)} \nu^{(n)} {\sigma^{(n)}}^2\big) +
  24 {b^{(n)}}^2 \big(315 \gamma_3^{(n)} -
\\
&&
{\nu^{(n)}}^2 \big(21 \beta^{(n)} + 42 c^{(n)} +
\nu^{(n)} {\sigma^{(n)}}^2\big)\big) +
  b^{(n)} \big(2520 \gamma_4^{(n)} \nu^{(n)} +
\sigma^{(n)} \big( 7560 \gamma_3^{(n)} -
\\
&&
      {\nu^{(n)}}^2 \big(294 \beta^{(n)} +
168 c^{(n)} + \nu^{(n)} {\sigma^{(n)}}^2\big)\big)\big)\big)/45360
\, ,
\end{eqnarray*}
\begin{eqnarray*}
\zeta_3^{(n+1)} =\zeta_3^{(n)} -
&&
\big( 2520 {\alpha^{(n)}}^2 b^{(n)} + 57 {b^{(n)}}^3 {\nu^{(n)}}^4 -
840 \alpha^{(n)} \nu^{(n)} \big( 3 c^{(n)} - {b^{(n)}}^2 \nu^{(n)}\big) +
  42 c^{(n)} {\nu^{(n)}}^3 \sigma^{(n)} -
12 {b^{(n)}}^2 \big(210 \gamma_1^{(n)} -
\\
&&
{\nu^{(n)}}^4 \sigma^{(n)}\big) -
  b^{(n)} \big(2520 \gamma_2^{(n)} \nu^{(n)} -
42 \beta^{(n)} {\nu^{(n)}}^3 + 336 c^{(n)} {\nu^{(n)}}^3 +
    2520 \gamma_1^{(n)} \sigma^{(n)} +
{\nu^{(n)}}^4 {\sigma^{(n)}}^2\big)\big)/15120
\, ,
\end{eqnarray*}
\begin{eqnarray*}
\zeta_4^{(n+1)} =\zeta_4^{(n)} +
&&
\big(5040 {\alpha^{(n)}}^2 b^{(n)} -
42 \alpha^{(n)} \nu^{(n)} \big( 120 c^{(n)} -
b^{(n)} \nu^{(n)} \big(36 b^{(n)} + \sigma^{(n)}\big)\big) +
  \nu^{(n)} \big(96 {b^{(n)}}^3 {\nu^{(n)}}^3 +
84 c^{(n)} {\nu^{(n)}}^2 \sigma^{(n)}
\\
&&
+
18 {b^{(n)}}^2 {\nu^{(n)}}^3 \sigma^{(n)} -
    b^{(n)} \big(5040 \gamma_2^{(n)} +
2520 \gamma_3^{(n)} - {\nu^{(n)}}^2 \big( 84 \beta^{(n)} - 672 c^{(n)} -
        5 \nu^{(n)} {\sigma^{(n)}}^2\big)\big)\big)\big)/15120
\, ,
\\ [-24pt]
\end{eqnarray*}
\begin{equation}
\end{equation}
\begin{eqnarray*}
\\ [-25pt]
\zeta_5^{(n+1)} =\zeta_5^{(n)} -
&&
\big( 2520 {\alpha^{(n)}}^2 b^{(n)} -
36 {b^{(n)}}^3 {\nu^{(n)}}^4 + 42 c^{(n)} {\nu^{(n)}}^3 \sigma^{(n)} +
  30 {b^{(n)}}^2 {\nu^{(n)}}^4 \sigma^{(n)} +
168 \alpha^{(n)} \nu^{(n)} \big( 15 c^{(n)} -
\\
&&
b^{(n)} \nu^{(n)} \big(6 b^{(n)} + \sigma^{(n)}\big)\big) -
  b^{(n)} \big(15120 \gamma_3^{(n)} \nu^{(n)} -
{\nu^{(n)}}^3 \big(252 \beta^{(n)} + 504 c^{(n)} +
\nu^{(n)} {\sigma^{(n)}}^2\big)\big)\big)/45360
\, ,
\end{eqnarray*}
\begin{eqnarray*}
\zeta_6^{(n+1)} =\zeta_6^{(n)} -
&&
\big( 630 {\alpha^{(n)}}^2 b^{(n)} + 27 {b^{(n)}}^3 {\nu^{(n)}}^4 -
21 c^{(n)} {\nu^{(n)}}^3 \sigma^{(n)} +
  9 {b^{(n)}}^2 {\nu^{(n)}}^4 \sigma^{(n)} -
63 \alpha^{(n)} \nu^{(n)} \big( 20 c^{(n)} -
\\
&&
b^{(n)} \nu^{(n)} \big(6 b^{(n)} + \sigma^{(n)}\big)\big) -
  b^{(n)} \big(1260 \gamma_2^{(n)} \nu^{(n)} +
{\nu^{(n)}}^3 \big(21 \beta^{(n)} + 252 c^{(n)} -
\nu^{(n)} {\sigma^{(n)}}^2\big)\big)\big)/3780
\, ,
\end{eqnarray*}
\begin{eqnarray*}
\zeta_7^{(n+1)} =\zeta_7^{(n)} -
&&
     b^{(n)} \nu^{(n)} \big( 2520 \gamma_1^{(n)} -
42 \alpha^{(n)} {\nu^{(n)}}^2 -
{\nu^{(n)}}^4 \big(6 b^{(n)} - \sigma^{(n)}\big)\big)/15120
\, ,
\end{eqnarray*}
\begin{eqnarray*}
\zeta_8^{(n+1)} =\zeta_8^{(n)} +
&&
\big(5040 b^{(n)} \gamma_1^{(n)} \nu^{(n)} -
42 c^{(n)} {\nu^{(n)}}^4 - 6 {b^{(n)}}^2 {\nu^{(n)}}^5 +
b^{(n)} {\nu^{(n)}}^5 \sigma^{(n)}\big)/15120
\, ,
\end{eqnarray*}
\begin{eqnarray*}
\zeta_9^{(n+1)} =\zeta_9^{(n)}
&&
-     {\nu^{(n)}}^3 \big(84 \alpha^{(n)} b^{(n)} -
\nu^{(n)} \big( 84 c^{(n)} - b^{(n)} \nu^{(n)} \big(12 b^{(n)} +
5 \sigma^{(n)}\big)\big)\big)/15120
\, ,
\end{eqnarray*}
\begin{eqnarray*}
\zeta_{10}^{(n+1)} =\zeta_{10}^{(n)} -     b^{(n)} {\nu^{(n)}}^6/15120
\, .
\end{eqnarray*}

\begin{multicols}{2}

All these relations as well as other symbolic expressions presented in
the work has been carried out using Mathematica 4.0 and Maple 6 packages
installed on the Silicon Graphics Origin 3800 workstation at Linz
University. The numerical calculations have also been performed there.

\end{multicols}

\vspace{12pt}

\begin{multicols}{2}

\end{multicols}

\end{document}